\documentclass[modern]{aastex631}
\usepackage{showyourwork}
\NewPageAfterKeywords
\usepackage{xspace}
\usepackage[utf8]{inputenc}
\usepackage[T1]{fontenc}
\usepackage{ulem}

\shorttitle{Astropy: Sustaining the Project and v5.0}
\shortauthors{The Astropy Collaboration}

\newcommand{\package}[1]{\texttt{#1}\xspace}
\newcommand{\github}{\package{GitHub}}
\newcommand{\python}{\package{Python}}
\newcommand{\astropy}{Astropy\xspace}
\newcommand{\astropypkg}{\package{astropy}}
\newcommand{\mission}[1]{\textit{#1}\xspace}

\renewcommand{\figurename}{Figure\xspace}

\renewcommand{\tablename}{Table\xspace}

\usepackage[colorinlistoftodos]{todonotes}

\usepackage{newunicodechar,graphicx}
\DeclareRobustCommand{\okina}{%
 \raisebox{\dimexpr\fontcharht\font`A-\height}{%
 \scalebox{0.8}{`}%
 }%
}
\newunicodechar{ʻ}{\okina}

\newcommand{\astropysubpkg}[1]{\href{http://docs.astropy.org/en/stable/#1/index.html}{\texttt{astropy.#1}}\xspace}
\newcommand{\astropyiosubpkg}[1]{\href{http://docs.astropy.org/en/stable/io/#1/index.html}{\texttt{astropy.io.#1}}\xspace}

\newcommand{\astropycosmology}{\astropysubpkg{cosmology}}
\newcommand{\astropycosmologyunits}{\href{https://docs.astropy.org/en/stable/cosmology/units.html}{\texttt{astropy.cosmology.units}}}
\newcommand{\astropyunits}{\astropysubpkg{units}}
\newcommand{\astropycoordinates}{\astropysubpkg{coordinates}}
\newcommand{\astropyconstants}{\astropysubpkg{constants}}
\newcommand{\astropytable}{\astropysubpkg{table}}
\newcommand{\astropytime}{\astropysubpkg{time}}
\newcommand{\astropymodeling}{\astropysubpkg{modeling}}

\newcommand{\astropyfits}{\astropyiosubpkg{fits}}

\newcommand{\astropyapi}[2]{\href{https://docs.astropy.org/en/stable/api/astropy.#1.html}{#2}}
\newcommand{\astropyapidoc}[2]{\astropyapi{#1}{\texttt{#2}\xspace}}

\newcommand{\astropySpectralCoord}{\astropyapidoc{coordinates.SpectralCoord}{SpectralCoord}}
\newcommand{\astropySkyCoord}{\astropyapidoc{coordinates.SkyCoord}{SkyCoord}}
\newcommand{\astropyICRS}{\astropyapidoc{coordinates.builtin_frames.ICRS}{ICRS}}
\newcommand{\astropyGalacticLSR}{\astropyapidoc{coordinates.builtin_frames.
GalacticLSR}{GalacticLSR}}
\newcommand{\astropyAltAz}{\astropyapidoc{coordinates.builtin_frames.AltAz}{AltAz}}
\newcommand{\astropyGalactocentric}{\astropyapidoc{coordinates.builtin_frames.Galactocentric}{Galactocentric}}
\newcommand{\astropyCosmology}{\astropyapidoc{cosmology.Cosmology}{Cosmology}}
\newcommand{\astropyFlatLambdaCDM}{\astropyapidoc{cosmology.FlatLambdaCDM}{FlatLambdaCDM}}
\newcommand{\astropyFitsColumn}{\astropyapidoc{io.fits.Column}{Column}}
\newcommand{\astropyModel}{\astropyapidoc{modeling.Model}{Model}}

\newcommand{\astropyTable}{\astropyapidoc{table.Table}{Table}}
\newcommand{\astropyQTable}{\astropyapidoc{table.QTable}{QTable}}
\newcommand{\astropyTime}{\astropyapidoc{time.Time}{Time}}
\newcommand{\astropyLeapSeconds}{\astropyapidoc{time.LeapSeconds}{LeapSeconds}}
\newcommand{\astropyDistribution}{\astropyapidoc{uncertainty.Distribution}{Distribution}}

\newcommand{\astropyQuantity}{\astropyapidoc{units.Quantity}{Quantity}}
\newcommand{\astropyMasked}{\astropyapidoc{utils.masked.Masked}{Masked}}
\newcommand{\astropyScienceState}{\astropyapidoc{utils.state.ScienceState}{ScienceState}}

\newcommand{\paperii}{\cite{astropy:2018}}

\newcommand{\astropyAPE}[1]{\href{https://github.com/astropy/astropy-APEs/blob/main/APE#1.rst}{APE #1}\xspace}
\newcommand{\gitHeader}{Git commit: 92722d7}

\defcitealias{astropy:2013}{Paper I}
\defcitealias{astropy:2018}{Paper II}

\begin{document}

\let\oldtoday\today
\renewcommand{\today}{\oldtoday;\qquad\textnormal{\gitHeader}}

\title{The Astropy Project: \\
       Sustaining and Growing a Community-oriented Open-source Project and
       the Latest Major Release (v5.0) of the Core Package}

\correspondingauthor{Astropy Coordination Committee}
\email{coordinators@astropy.org}

\author{Astropy~Collaboration}
\noaffiliation
{\let\thefootnote\relax\footnote{{
  The author list has two parts: the authors that made significant contributions
  to the writing and/or coordination of the paper (in order of contribution),
  followed by maintainers of and contributors to the Astropy Project (in
  alphabetical order). \textbf{The position in the author list does not
  correspond to contributions to the Astropy Project as a whole.} A more
  complete list of contributors to the core package can be found in the
  \href{https://github.com/astropy/astropy/graphs/contributors}{package
  repository}, and at the \href{http://www.astropy.org/team.html}{Astropy team
  webpage}.
}}}

\newcommand{\affstsci}{
    Space Telescope Science Institute, 3700 San Martin Drive, Baltimore, MD 21218, USA
}
\newcommand{\affcca}{
    Center for Computational Astrophysics, Flatiron Institute, 162 5th Ave, New York, NY 10010, USA
}
\newcommand{\affcfa}{
    Center for Astrophysics | Harvard \& Smithsonian, 60 Garden Street, Cambridge, MA, 02138, USA
}
\newcommand{\affbaeri}{
    Bay Area Environmental Research Institute, P.O. Box 25, Moffett Field, CA 94035, USA
}
\newcommand{\affprinceton}{
    Department of Astrophysical Sciences, 4 Ivy Lane, Princeton University, Princeton, NJ 08544, USA
}
\newcommand{\affaperio}{
    Aperio Software Ltd., Insight House, Riverside Business Park, Stoney Common Road, Stansted, Essex, CM24 8PL, UK
}
\newcommand{\afftorontoastro}{
    David A. Dunlap Department of Astronomy \& Astrophysics, University of Toronto, 50 St. George Street, Toronto ON M5S 3H4, Canada
}
\newcommand{\affipac}{
    IPAC, MC 100-22, Caltech, 1200E. California Blvd. Pasadena, CA 91125
}
\newcommand{\affcoco}{
    Astropy Coordination Committee
}
\newcommand{\affsteward}{
    Steward Observatory, University of Arizona, 933 North Cherry Avenue, Tucson,
    AZ 85721-0065, USA
}
\newcommand{\affsarao}{
    South African Radio Astronomy Observatory, 2 Fir Street, Black River Park,
    Observatory 7925, South Africa
}
\newcommand{\affindy}{
    Independent Astropy Contributor
}

% Paper coordinators, ordered by contributions:
\author[0000-0003-0872-7098]{Adrian~M.~Price-Whelan}
\altaffiliation{\affcoco}
\affiliation{\affcca}

\author[0000-0003-0079-4114]{Pey~Lian~Lim}
\affiliation{\affstsci}

\author[0000-0003-1714-7415]{Nicholas~Earl}
\affiliation{Department of Astronomy, University of Illinois at Urbana-Champaign, 1002 W. Green St., Urbana, IL 61801 USA}

% ----------------------------------------------------------------------------
% Paper contributors, ordered by contributions:
\author[0000-0003-3954-3291]{Nathaniel~Starkman}
\affiliation{\afftorontoastro}

\author[0000-0002-7908-9284]{Larry~Bradley}
\affiliation{\affstsci}

\author[0000-0003-4401-0430]{David~L.~Shupe}
\affiliation{\affipac}

\author[0000-0002-7626-506X]{Aarya~A.~Patil}
\affiliation{\afftorontoastro}
\affiliation{Dunlap Institute for Astronomy \& Astrophysics, University of Toronto, 50 St George Street, Toronto, ON M5S 3H4, Canada}

\author[0000-0002-5466-3817]{Lia~Corrales}
\affiliation{University of Michigan Astronomy, 1085 S. University, Ann Arbor MI 48109, United States}

\author[0000-0002-9314-960X]{C.~E.~Brasseur}
\affiliation{School of Physics and Astronomy, University of St Andrews, North Haugh, St Andrews KY16 9SS, Scotland}

\author[0000-0001-7993-8189]{Maximilian~N\"{o}the}
\affiliation{Astroparticle Physics, TU Dortmund, Otto-Hahn-Str. 4a, 44227 Dortmund, Germany}

\author[0000-0003-4568-7005]{Axel~Donath}
\affiliation{\affcfa}

\author[0000-0002-9599-310X]{Erik~Tollerud}
\altaffiliation{\affcoco}
\affiliation{\affstsci}

\author[0000-0003-2528-3409]{Brett~M.~Morris}
\affiliation{Center for Space and Habitability, University of Bern, Gesellschaftsstrasse 6, 3012 Bern, Switzerland}

\author[0000-0001-6431-9633]{Adam~Ginsburg}
\affiliation{Department of Astronomy, University of Florida, P.O. Box 112055, Gainesville, FL, USA}

\author{Eero~Vaher}
\affiliation{Lund Observatory, Department of Astronomy and Theoretical Physics, Lund University, Box 43, 22100 Lund, Sweden}

\author{Benjamin~A.~Weaver}
\affiliation{NSF's National Optical-Infrared Astronomy Research Laboratory, 950 North Cherry Avenue, Tucson, AZ 85719, United States}

\author[0000-0001-6637-6922]{James~Tocknell}
\affiliation{School of Mathematical and Physical Sciences, Macquarie University, Sydney, NSW 2109, Australia}
\affiliation{Astronomy, Astrophysics and Astrophotonics Research Centre, Macquarie University, Sydney, NSW 2109, Australia}
\affiliation{Australian Astronomical Optics, Macquarie University, Sydney, NSW 2109, Australia}

\author[0000-0001-5976-4492]{William~Jamieson}
\affiliation{\affstsci}

\author[0000-0002-5830-8505]{Marten~H.~van~Kerkwijk}
\affiliation{\afftorontoastro}

\author[0000-0002-8642-1329]{Thomas~P.~Robitaille}
\affiliation{\affaperio}

\author{Bruce~Merry}
\affiliation{\affsarao}

\author{Matteo~Bachetti}
\affiliation{Istituto Nazionale di Astrofisica/Osservatorio Astronomico di Cagliari, via della Scienza 5, I-09047 Selargius (CA), Italy}

\author[0000-0003-4243-2840]{H.~Moritz~G\"{u}nther}
\altaffiliation{\affcoco}
\affiliation{MIT Kavli Institute for Astrophysics and Space Research, 77 Massachusetts Avenue, Cambridge, MA 02139, USA}

\collaboration{23}{(Paper Authors)}

% ----------------------------------------------------------------------------
% All others, alphabetical:
\author{Thomas~L.~Aldcroft}
\affiliation{\affcfa}

\author[0000-0003-0353-9741]{Jaime~A.~Alvarado-Montes}
\affiliation{School of Mathematical and Physical Sciences, Macquarie University, Sydney, NSW 2109, Australia}
\affiliation{Astronomy, Astrophysics and Astrophotonics Research Centre, Macquarie University, Sydney, NSW 2109, Australia}

\author[0000-0003-0638-3340]{Anne~M.~Archibald}
\affiliation{Newcastle University}

\author[0000-0002-8585-4544]{Attila~B\'odi}
\affiliation{Konkoly Observatory, Research Centre for Astronomy and Earth Sciences (ELKH), H-1121 Budapest, Konkoly Thege Mikl\'os \'ut 15-17, Hungary}
\affiliation{MTA CSFK Lend\"ulet Near-Field Cosmology Research Group}

\author[0000-0002-0870-4665]{Shreyas~Bapat}
\affiliation{School of Computing and Electrical Engineering, Indian Institute of Technology Mandi, Kamand, Mandi (H.P.), 175005, India}

\author[0000-0002-3306-3484]{Geert~Barentsen}
\affiliation{\affbaeri}

\author[0000-0001-7699-3983]{Juanjo~Baz{\'a}n}
\affiliation{CIEMAT, Astroparticle physics, Madrid, Spain}

\author{Manish~Biswas}
\affiliation{Savitribai Phule Pune University, Pune, Maharashtra, India}

\author[0000-0003-0946-6176]{M\'ed\'eric~Boquien}
\affiliation{Centro de Astronomía (CITEVA), Universidad de Antofagasta, Avenida Angamos 601, Antofagasta, Chile}

\author[0000-0003-4428-7835]{D.~J.~Burke}
\affiliation{\affcfa}

% ...

\collaboration{113}{(Astropy Project Contributors)}

\begin{abstract}
The \astropy Project supports and fosters the development of open-source and openly developed
\python packages that provide commonly needed functionality to the astronomical
community.
A key element of the \astropy Project is the core package \astropypkg, which serves as the
foundation for more specialized projects and packages.
In this article, we summarize key features in the core package as of the recent major
release, version 5.0, and provide major updates for the Project.
We then discuss supporting a broader ecosystem of interoperable packages,
including connections with several astronomical observatories and missions.
We also revisit the future outlook of the \astropy Project and the current
status of Learn Astropy.
We conclude by raising and discussing the current and future challenges facing the Project.
\end{abstract}

% \keywords{%
%     Astrophysics - Instrumentation and Methods for Astrophysics
%     ---
%     methods: data analysis
%     ---
%     methods: miscellaneous
% }

\section{Introduction} \label{sec:intro}

% \secauthor{Adrian Price-Whelan}

\begin{figure}[t!]
    \begin{centering}
      \includegraphics[width=\textwidth]{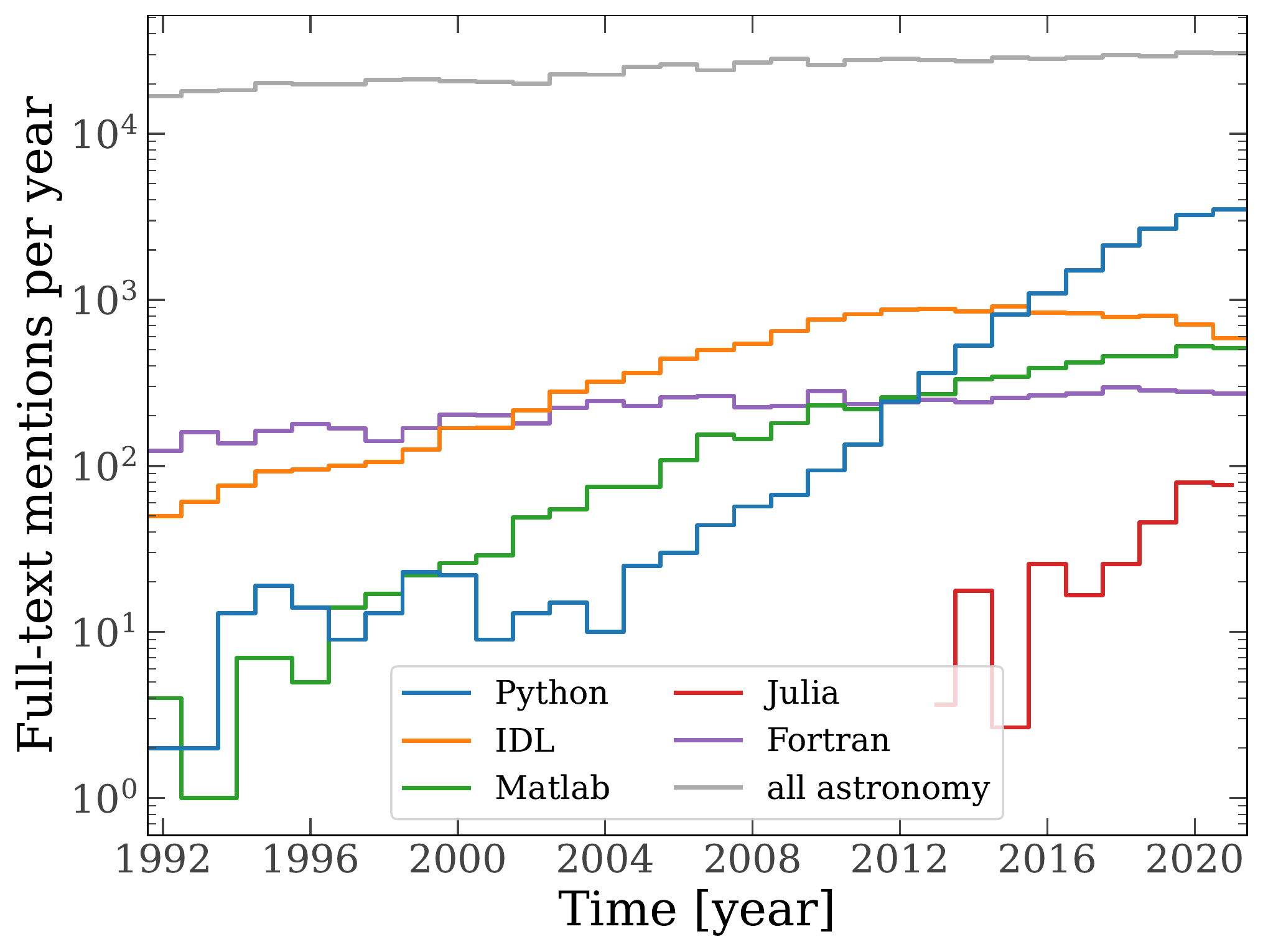}
        \caption{
            Yearly full-text mentions of programming languages (indicated in the
            figure legend) in refereed publications in the astronomical
            literature database in the Astrophysics Data System (ADS).
            \python has rapidly become the dominant programming language
            mentioned in refereed articles over the last 10 years.
        }
        \label{fig:python-mentions}
    \end{centering}
\end{figure}

The \python programming language is a high-level, interpreted (as opposed to
compiled) programming language that has become an industry standard across many
computational domains, technological sectors, and fields of research.
This recent and rapid adoption of \python stems from the fact that it enables
scalable, time- and energy-efficient code execution \citep[e.g.,][]{Augier:2021}
with a focus on code readability, ease of use, and interoperability with other
languages.
Over the last decade, \python has grown enormously in popularity to become a
dominant programming language, especially in the astronomical and broader scientific
communities.
For example, Figure~\ref{fig:python-mentions} shows the number of yearly
full-text mentions of \python as compared to a selection of other programming languages
in refereed articles in the astronomical literature, demonstrating its nearly
exponential growth in popularity.
The adoption of \python by astronomy researchers, students, observatories, and
technical staff combined with an associated increase in awareness and interest
about open-source software tools is contributing to a paradigm shift in the way
research is done, data is analyzed, and results are shared in astronomy and
beyond.

One of the factors that has led to its rapid ascent in popularity in scientific
contexts is the combination of volunteer-driven and professionally-supported
effort behind developing community-oriented open-source software tools and
fostering communities of users and developers that have grown around these
efforts.
Today, a broad and feature-diverse ``ecosystem'' of packages exists in the
\python scientific computing landscape: roughly ordered from general-use to
domain-specific, this landscape now includes packages that provide core
numerical analysis functionality like \package{numpy} \citep{numpy:nature} and
\package{scipy} \citep{scipy}, visualization frameworks like
\package{matplotlib} \citep{matplotlib}, machine learning and data analysis
packages like \package{tensorflow} \citep{tensorflow}, \package{pymc3}
\citep{Salvatier:2016}, and \package{emcee} \citep{emcee}, domain-specific
libraries like \package{yt} \citep{yt:2011}, \package{plasmapy}
\citep{plasmapy}, \package{sunpy} \citep{sunpy:apj}, \package{Biopython}
\citep{biopython}, and \package{sympy} \citep{sympy} (to name a few in each
category).
The \astropypkg \citep{astropy:2013, astropy:2018} core package began in this
vein, as an effort to consolidate the development of commonly used functionality
needed to perform astronomical research into a community-developed \python
package.

The \astropypkg core package was one of the first large, open-source \python
packages developed for astronomy and provides, among other things, software
functionality for reading and writing astronomy-specific data formats (e.g.,
FITS), transforming and representing astronomical coordinates, and representing
and propagating physical units in code.
An early description of the core functionality in \astropypkg can be found in
the first \astropy paper \citep{astropy:2013} or in detail in the
\href{https://docs.astropy.org/}{core package documentation}.
The codebase in the \astropypkg core package is now largely stable, in that the
software interface does not change without sufficient and significant
motivation, and the addition of new features into the core package has slowed
as compared to the first years of its development.
This is largely driven by the fact that the core package now represents just one
piece of the broader astronomy \python context, and new feature development
is now happening in more specialized packages that are expanding the
capabilities of the \astropy ecosystem by building on top of the foundations
laid by the \astropypkg core package.
Because of this natural expansion, the name \astropy has grown in scope beyond a
single \python library to become ``the Astropy Project.''

The \astropy Project is a community effort that represents the union of the
\astropypkg core package, the ecosystem of astronomy-specific software tools
that are interoperable with \astropypkg (Astropy Coordinated and Affiliated Packages),
\emph{and} the community of users, developers, and maintainers that participate
in Astropy efforts.
However, there is no single institution responsible for managing the Astropy Project,
for funding or maintaining its development, or sustaining it into the future:
The Project is maintained and coordinated by both volunteers and contributors
who are primarily paid for other reasons (i.e., Astropy work is just a means to an end).
While new Astropy-affiliated packages are being developed that expand upon the
core functionality in the \astropypkg package, representing a natural expansion
of the Astropy Project ecosystem, the needs of and challenges faced by the
Project are evolving.
In particular, fostering the development of the community and its expansion has
been an additional key goal of the Astropy Project over the last several years.

In this Article, we briefly describe recent key updates in the \astropypkg core
package since the last Astropy paper (``Paper II''; \citealt{astropy:2018}),
major updates in the governance, contributor base, and funding of the Project,
and discuss some of the future plans and challenges faced by the Astropy
Project.

\section{Major Updates to the Astropy Core Package} \label{sec:core-updates}

\subsection{New Long-term Support (LTS) Version: v5.0} \label{sec:core-v50}

% \secauthor{Tom Robitaille}

Major versions of the core package --- that is, versions that add and/or modify
functionality --- are released approximately every six months, and are then
maintained with releases that fix issues until the next major version is
released. However, every two years a major release is designated as a long-term
support (LTS) release, which continues to be maintained for at least two years
\citep{ape2}. The motivation for LTS releases is to provide longer-term stable
versions of \astropypkg that users who require a high level of stability can
make use of if they do not always need the latest features (this includes but
is not limited to telescope operations and data reduction pipelines). The v5.0
release of the core package, replacing v4.0, was designated as LTS at the end of 2021;
it will be maintained until the end of 2023.

\subsection{Highlighted Feature Development} \label{sec:core-features}

% \secauthor{Nathaniel Starkman, Marten van Kerkwijk}

The \astropypkg core package is mature and stable, in that many features have
been part of the core package for years with purposefully few changes to the
software interface. This maturity and stability allows for the broader
astronomy \python community to rely upon the \astropypkg core package and build
specialized packages within the Astropy ecosystem. Even many non-astronomical
\python libraries have come to rely on \astropypkg. The relationship between
\astropy and the broader \python community is reciprocal, with the evolving needs
of the community and the maturation of the broader scientific \python ecosystem
driving much of the development of the \astropypkg core package.

Within \astropypkg, development may be roughly split into several
categories: new features, intra- and inter-package interoperability,
improvements to numerical precision, accuracy, and reproducibility, performance
and documentation enhancements, and bug fixes. We discuss some of these in the subsections below.

Additionally, some features have been moved from the core package, either
because they are of more general use outside of \astropypkg, or because they are
too specialized and belong in an Astropy-affiliated package. An example of the
former is the copy of the IAU Standards Of Fundamental Astronomy (SOFA) software
Collection\footnote{\url{http://www.iausofa.org}} \citep{sofa} that \astropy
carried, as well as the associated \python wrappers. Since this is basic
infrastructure, with a release cycle set by SOFA, it made more sense to create
separate packages that serve as \astropypkg dependencies: \package{ERFA}
\citep{erfa} and \package{PyERFA} \citep{pyerfa}. An example of something that
was too specialized was a Virtual Observatory subpackage; the Simple
Application Message Protocol (\astropysubpkg{samp}) is kept in the core package
for now, but other parts more properly belonged in
\package{astroquery} (see Section~\ref{sec:astroquery}).

\subsubsection{New and Planned Features} \label{sec:core-features-new}

New features present a particular difficulty for a mature package such as
\astropypkg, because it is challenging to know {\em a priori} what the best
interface will be, and hard to address all possible use cases in one round of
development, even within \astropypkg\ itself. New features come in various forms,
from new submodules providing wholly new capabilities, to larger additions to
existing submodules, to large rewrites of the interface of modules. This range
is spanned by five main new features that entered the \astropypkg core since the
previous summary \citepalias{astropy:2018}. In our descriptions of each, we
include how we hope the new feature will evolve. Ultimately, in a user-driven
project like \astropy, development depends primarily on the priorities of
contributors.

\paragraph{Uncertainties and Distributions} In scientific analysis,
measurement values with units and uncertainties are essential.
\astropyunits provides the ability to associate numbers with units, and
propagate these correctly, using the \astropyQuantity\ class. However, the
ability to associate and propagate uncertainties is relatively limited: in
\astropysubpkg{nddata}, there are options to associate errors with data arrays,
but covariances are not tracked.

Error propagation is difficult, and often is best approximated using Monte Carlo
methods. The new \astropysubpkg{uncertainties} subpackage is our first step
towards enabling seamless error propagation. The package allows one to
generate, for each variable, randomly drawn samples in a \astropyDistribution,
and then propagate these by passing them through the normal analysis, producing
a \astropyDistribution\ of final results that can be inspected. This
\astropyDistribution\ can be any type of array, including a \astropyQuantity.
The goal is to ensure \astropyDistribution\ can be seamlessly used to
instantiate other \astropypkg classes as well, such as \astropySkyCoord,
\astropyTime, and \astropyCosmology.

A plan for future development of \astropysubpkg{uncertainties} is to
support error propagation: tracking covariances for the case that uncertainties
are normally distributed rather than approximated by Monte-Carlo means. One
implementation problem to be solved is when to {\em stop} tracking covariances.
For instance, a simple operation such as subtracting the mean from $N$ data
points implies that all data points are now covariant with each other, i.e., one
has to carry $N\times N$ covariances. For a large image, that becomes unnecessary
and computationally expensive.

\paragraph{Masked quantities}  It can be useful to mask bad data, either by marking
bad data values with a flag or by replacing them with a special
value, e.g., ``Not a Number'' (NaN). In \astropypkg, both approaches are
used: a flag for masked columns in \astropyTable, mask flags and bitmaps
for N-dimensional data in \astropysubpkg{nddata}, and replacing elements with
NaN in \astropyTime. However, masked \astropyQuantity\ support is currently very limited,
hindering usage of masks in a \astropyQTable\ (in which \astropyQuantity\ is used
for all columns with units).

Furthermore, the \texttt{MaskedArray} class from \package{numpy} only works
well with plain data arrays, hiding other metadata and attributes, for example
the \texttt{unit} of a \astropyQuantity. Consequently, a new \astropyMasked\
class was designed, based on a framework very similar to that of
\astropyDistribution, making it easy to create masked instances of other
classes. In particular, \astropyMasked\ can be used to create masked quantities
for \astropyQTable, enabling I/O of masked data from different file types.

The masked quantities work largely without any changes in higher-level objects
such as \astropySkyCoord, but more work is needed to better integrate it with
the rest of \astropypkg, e.g., using masked arrays in \astropyTime\ instead of NaN.

\paragraph{Time series}  From sampling a continuous variable at fixed times to
counting events binned into time windows, many different areas of astrophysics
require the manipulation of time series. The new \astropysubpkg{timeseries}
subpackage extends the \astropyQTable\ class to support tables of data as a
function of time, where the data can either represent samples or averages over
particular time bins. The new classes offer a number of special methods to
manipulate time series (e.g., folding and resampling) and to read different data
formats (e.g., \mission{Kepler} light curves).

Also part of \astropysubpkg{timeseries} are common analysis routines, including
\href{https://docs.astropy.org/en/stable/timeseries/lombscargle.html}{Lomb-Scargle}
and
\href{https://docs.astropy.org/en/stable/timeseries/bls.html}{box-least-squares}
periodograms.

\paragraph{Spectral Coordinates} Measurements are often taken at specific
``spectral coordinates,'' be they frequencies, wavelengths, or photon energies.
\astropyQuantity\ can represent these and convert between them via dedicated
equivalencies. The new \astropySpectralCoord\ builds on \astropyQuantity\ by
providing a more straightforward interface: baking in these equivalencies and
those for Doppler velocities. Furthermore, \astropySpectralCoord\ can be made
aware of the observer and target reference frames, allowing transformation from
telescope-centric (or topocentric) frames to Barycentric or Local Standard of
Rest (LSRK and LSRD) velocity frames.

\paragraph{A High-Level Interface to World Coordinate Systems}
Astronomical data are often provided alongside information about the
correspondence between ``real-world'' and pixel coordinates. This mapping is the
essence of the World Coordinate System (WCS) concept. From its inception,
the \astropysubpkg{wcs} subpackage allowed access to WCS information provided
in, e.g., FITS files, but it became clear that other WCS standards and
representations had to be supported for new missions and observatories (e.g.,
JWST and the Rubin Observatory). To harmonize these
needs, a new high-level interface was created based on a formal design first
proposed in an \astropy\ Proposal for Enhancement (\astropyAPE{14}). The hope is
that by having a formal design, other packages implementing WCS objects can
straightforwardly modify their classes to conform to the new interface or build
thin wrappers that conform. The implementation in \astropysubpkg{wcs} interacts
well with other \astropypkg\ objects such as \astropySkyCoord\ and \astropyTime.
Future work includes better integration of this interface with the rest of
the Project.

\subsubsection{Interoperability} \label{sec:core-features-interoperability}

Because \astropy is modular and situated at the nexus between scientific
computing and astronomy, interoperability is an important focus for \astropy,
both within the various \astropypkg subpackages and with other \python
libraries. We aim for a seamless user experience, where different code blocks
would ``just work'' when put together. Here, we describe a few particular
efforts towards this goal.

\paragraph{\package{numpy} on Units}

\astropyQuantity\ is a backbone of \astropy, leveraging the power of
\package{numpy} and adding units. Previously, many \package{numpy} functions
would strip a \astropyQuantity\ of its units (or fail outright), limiting
\astropyQuantity's potential. Now, advances in \package{numpy} function
overloading (e.g.,
\href{https://numpy.org/neps/nep-0018-array-function-protocol.html}{NEP 18})
mean \astropyQuantity\ works with almost all \package{numpy} (v1.17+)
functions.

In the remaining gaps, the \package{numpy} and \astropyQuantity\
interoperability efforts are ongoing. For some functions, such as in
\href{https://numpy.org/doc/stable/user/basics.rec.html#module-numpy.lib.recfunctions}{\package{numpy}'s
module} for manipulating structured arrays, compatibility only requires
extending the existing \package{numpy}-\astropypkg bridge frameworks. Community
interest, in the form of a Feature Request (or better yet Pull Request) would be
sufficient to see this compatibility completed. For a few remaining functions,
discussed further in the
\href{https://docs.astropy.org/en/stable/known_issues.html#known-deficiencies}{\astropy
documentation}, the \package{numpy} framework does not yet allow for full
interoperability with \astropy. A goal of \astropy and the scientific \python
community is to enhance and implement the frameworks, allowing \astropyQuantity\
to propagate units seamlessly across the whole ecosystem of \package{numpy}-like
projects.

\paragraph{\astropypkg on Units}

Units were also integrated further within \astropypkg. For an already-defined
unit-less \astropyModel, e.g., those imported from another library,
\astropymodeling can now coerce units. A new module – \astropycosmologyunits –
has been added to the cosmology subpackage for defining and collecting
cosmological units and equivalencies. \astropycosmologyunits\ has a unit and
equivalencies for \texttt{littleh}, and a new unit, \texttt{redshift}, for
factors of cosmological redshift. \texttt{redshift} has a number of
equivalencies for converting between cosmological distance measures, e.g., CMB
temperature or comoving distance.

\paragraph{Table Column Types}

Within \astropypkg and \python, there are numerous different ways to represent
and store array-valued data. Some differences are historical: \astropytable and
\astropyapidoc{io.fits.BinTableHDU}{io.fits.BinTableHDU} are both for holding
tabular data but the latter was developed for and remains largely specific to
FITS. Some differences are computational: \package{numpy}, \package{cupy}, and
\package{dask} arrays have almost identical APIs, but are optimized for
different use cases. Lastly, some differences are inherent: \astropyQuantity,
\astropyTime, and \astropySkyCoord\ represent fundamentally different types of
objects.

\astropytable is \astropypkg's most general container for array-valued data.
Rather than requiring component data be converted to a
\astropyapidoc{table.Column}{table.Column}, a well-defined protocol has been
developed for any array-valued data to be used in a \astropytable, without
having to convert the data. The protocol for mixing column types, which \astropypkg
terms ``mixin''-columns, provides the same familiar table API, yet loses none
the object's data or attributes. Note that ``mixin'' here is not implemented
in the commonly used Python scheme of multiple inheritance.
Rather, it is a composition protocol that makes it
possible to store \astropy native objects -- including \astropyTime,
\astropyQuantity, and \astropySkyCoord\ -- within a \astropyTable\ and write
these to various file formats, such as FITS. For objects not already covered by
the \astropypkg mixin-protocol, functions can be registered with \astropyTable\ to
enable interoperability. As an example, \astropypkg already implements these
mixin functions to integrate \package{dask} arrays with \astropyTable, allowing
cloud-stored or cluster-scale data to be used as a column in a \astropyTable.

\paragraph{Astropy FITS in Time}

The FITS standard was extended to rigorously describe time coordinates in the
WCS framework \citep{FITS-Time:2015}. Compared to other types of coordinates in
WCS, time requires more metadata: format, scale, position, reference, etc. This
metadata had to be manually specified and the nuances understood by the user.
\astropyfits could read this data as a standard \astropyFitsColumn, but would
not interpret the time-related metadata and attributes. Through the support of
the
\href{https://summerofcode.withgoogle.com/archive/2017/projects/4778482366152704}{Google
Summer of Code} 2017 program,\footnote{To learn more about this project, see the
final report
\href{https://aaryapatil.wordpress.com/2017/08/28/a-mixin-protocol-for-seamless-interoperability/}{A
mixin protocol for seamless interoperability}.} the \astropyfits package now
interprets time data correctly, using \astropyTime\ as a column. For backward
compatibility, this feature may be turned off.

\paragraph{Persistent Storage}

\astropy's efforts to increase support for column types in a table is mirrored
by efforts to expand storage format options. New formats have been added, and
existing formats updated to support more column types.

\astropy now supports reading and writing tables in the American Astronomical
Society Journals' Machine-Readable Table (\href{https://docs.astropy.org/en/stable/io/ascii/write.html#cds-mrt-format}{MRT}) format. This ASCII format has
long been missing and was added to \astropyTable\ I/O in a Google Summer of Code 2021 project.
In addition, \astropyTable\ may also read from and write
to \href{https://asdf-standard.readthedocs.io/en/latest/}{ASDF},\footnote{Starting in v5.1,
direct \astropypkg support for ASDF has been deprecated in favor of moving it
to the new \package{asdf-astropy} \citep{asdf-astropy} package to accommodate
the differing release schedules between ASDF and \astropypkg.}
\href{https://docs.astropy.org/en/stable/io/unified.html#table-io-parquet}{Parquet},
and \href{https://wwwastro.msfc.nasa.gov/qdp/}{QDP} formats.

The \href{https://github.com/astropy/astropy-APEs/blob/main/APE6.rst}{ECSV
standard} has been updated to v1.0, adding support of three additional data
subtypes: multidimensional column data (both masked and unmasked)
with fixed dimensions in all table cells, multidimensional
column data with variable-dimension arrays similar to FITS variable-length
arrays, and object-type columns with simple \python objects.

As mentioned previously, \package{dask} arrays may be used as a column in
\astropyTable. Now, \package{dask} may also be a data array in
\astropyapidoc{io.fits.PrimaryHDU}{FITS HDU} and only computed while written,
avoiding excessive memory use. Furthermore, \astropyTable\ can be appended to an
existing FITS file, where \package{dask} columns can interoperate seamlessly
between them.

\paragraph{Unified I/O architecture}

\astropyiosubpkg{registry} is a powerful way to define input and output (I/O)
functions for \astropypkg objects, such as reading from or writing to a file,
following a given standard. \astropypkg uses this internally for the I/O methods in
\astropyTable\ and \astropyCosmology. Users can also register custom I/O to
extend the options on these classes.

As of \astropypkg v5.0, the I/O registry subpackage has been generalized to enable a
number of new use cases; for instance, the class-based architecture allows
streamlining of the creation of custom registries while minimizing code duplication.
One application is in \astropycosmology,
which has two different \textit{kinds} of registries for \astropyCosmology: one
for reading and writing files, and another for converting between \python objects.

\subsection{Numerical Precision, Accuracy, \& Reproducibility}
\label{sec:core-features-precision_accuracy_repreducibility}

An important focus of \astropypkg has been to increase the numerical precision and accuracy
of its functionality, while, where appropriate, making sure results obtained
using older versions of \astropypkg are reproducible. Within each section below,
we include in our descriptions how old results may be reproduced. The primary
means are with options in configuration files and settings on runtime configuration
objects called \astropyScienceState. Notable improvements to \astropypkg have
been made to \astropytime, \astropyconstants, \astropycoordinates, and
\astropycosmology, and we describe each in turn.

\paragraph{Time}
In astronomy, time accuracy down to the (nano)second is frequently important.
Whether for planning observations or crunching pulsar or VLBI data, missing
seconds meaningfully impact results. In past versions of \astropypkg, when using
\astropyTime, leap second tables had to be manually updated. Now, \astropypkg
automatically updates internal time data files to ensure that the correct leap
second adjustment is always used. For reproducibility with older code,
\astropyLeapSeconds\ may also be manually applied.

\paragraph{Constants}

Measurements of physical constants (and the units defined from them) improve
over time. Periodically the standardized systems of units and constants are
updated to reflect these improvements. For instance, in 2019 the SI system was
redefined \citep{NIST2019}, with an accompanying update to the physical
constants in
\href{https://codata.org/initiatives/data-science-and-stewardship/fundamental-physical-constants/}{\texttt{SI/CODATA
2018}}. \astropyconstants now defaults to using the \texttt{SI/CODATA 2018}
values, with the units in \astropyunits based on these constants. Most of
\astropy and affiliated packages build upon \astropyunits, so this update
affects the entire \astropy ecosystem.

For reproducibility, \astropypkg allows the constants' (and therefore units')
definitions to be rolled back to prior values, e.g., to \texttt{SI/CODATA
2014}. For work sensitive to the values of the fundamental constants, we
recommend including an \astropypkg configuration file (i.e., \texttt{astropy.cfg})
with the work, which specifies the set of constants used.

\paragraph{Cosmology}

\astropycosmology contains classes for representing cosmological models. Bundled
with the classes are commonly used \astropyCosmology\ realizations, e.g.,
best-fit measurements from
\href{https://lambda.gsfc.nasa.gov/product/map/current/}{WMAP}
\citep{WMAP2003} and
\href{https://www.nasa.gov/mission_pages/planck}{Planck}
\citep{PlanckMission:2006}.

When these important missions publish new measurements, \astropycosmology is
updated to include these results as realizations of the appropriate class
(generally \astropyFlatLambdaCDM). The models used by WMAP and Planck do not
always exactly correspond to \astropycosmology classes. For instance, the Planck
2018 results \citep{Planck2018VI:2020} include massive neutrinos in
$\Omega_{matter,0}$, while in \astropyFlatLambdaCDM\ this mass contribution is
stored in a separate parameter. Consequently, while some parameter values appear
different from the source paper, the \astropyCosmology\ realizations correctly
reproduce the cosmological models.

Since \citet{astropy:2018}, the following cosmology realizations have been
added:
\href{http://docs.astropy.org/en/stable/api/astropy.cosmology.WMAP1.html}{WMAP First Year}
\citep{WMAP1Year:2003};
\href{http://docs.astropy.org/en/stable/api/astropy.cosmology.WMAP3.html}{WMAP
Three Year} \citep{WMAP3Year:2007};
\href{http://docs.astropy.org/en/stable/api/astropy.cosmology.Planck15.html}{Planck 2015}
\citep{Planck2015XIII:2016}; and the
\href{http://docs.astropy.org/en/stable/api/astropy.cosmology.Planck18.html}{Planck
2018} best-fit cosmological parameters \citep{Planck2018VI:2020}.

\astropypkg provides a
\href{http://docs.astropy.org/en/stable/api/astropy.cosmology.default_cosmology.html}{configurable
default cosmology}, which is used in calculations done in a cosmological
context. The default \astropyCosmology\ has been updated to the Planck 2018
parameters. For reproducibility, this cosmology may be set to old defaults, such
as the WMAP Three Year values. The default cosmology is dynamically configurable
using \astropyScienceState, meaning a set of calculations may be run using
different assumed cosmologies, and results compared between the two.

\paragraph{Coordinates}

Being able to describe the coordinates of objects is
fundamental across many knowledge domains. \astropycoordinates allows one to
work with low-level positional and velocity data all the way to high-level
coordinate objects with reference frames, atmospheric information, etc. A
central feature of \astropycoordinates is the ability to transform data between
reference frames, e.g., \astropyICRS\ \citep{ICRS:1997} to \astropyGalacticLSR\
\citep{GalacticLSR:2010}. However, there have been some historical limitations
for spatially proximate transformations, impacting the usefulness of
\astropycoordinates for ground-based telescopes and astrometry within the
solar system. Transformations in the \astropyAltAz\ frame were reasonably
precise for very distant objects, but wrong by up to several arcseconds for,
e.g., the location of the moon. Now these transformations are much more precise,
down to the milliarcsecond level. Similar precision improvements were made to
the
\astropyapi{coordinates.builtin_frames.HADec}{Hour Angle-Declination}
frame transformations. Additionally,
\astropyapi{coordinates.builtin_frames.BaseEclipticFrame}{Ecliptic frames}
and associated transformations have been updated to correctly reflect the ``true''
and ``mean'' terminology.

For Galactic astronomers, the \astropyGalactocentric\
\astropyapi{coordinates.galactocentric_frame_defaults}{frame defaults}
have been updated to include more recent measurements (tabulated in the
\href{https://docs.astropy.org/en/stable/api/astropy.coordinates.galactocentric_frame_defaults.html#astropy.coordinates.galactocentric_frame_defaults.references}{\texttt{frame\_defaults.references}}
attribute). For reproducibility, old definitions are still available.

\subsection{Performance} \label{sec:core-features-performance}

In some cases, feature enhancements and corner-case handling were added
at the cost of performance.
Hence, \github pull requests that improve performance are welcomed, and were
even the main goal for v3.1 release. For most subpackages, performance was
improved in the \python\ code, but for some, the most time-critical pieces were
rewritten in~C and wrapped with \package{Cython} (\citealt{cython}; e.g., for
convolution of images, sigma-clipping, and converting time strings to binary).
Furthermore, a particular effort was made to ensure \astropypkg\ is thread-safe,
so that it can be used on supercomputing clusters.

Performance in \astropypkg is addressed based on feedback and requests, and
through contributions from community members: if you find that parts of
\astropypkg are not performant, we encourage you to open a \github issue to
start a discussion with the \astropypkg core maintainers.

\section{Major Updates in the Astropy Project} \label{sec:project-updates}

\subsection{Project governance} \label{sec:project-governance}

% \secauthor{Erik Tollerud}

%Briefly summarize new procedures and governance structure, new CoCo, election
%process overview, etc.

As part of the process of developing \astropypkg into a long-term sustainable
product, and to improve transparency and accountability, the Project agreed to
write down and formalize our governance structure (partly supported by explicit
funding for this purpose; see Section~\ref{sec:project-funding}). At the 2019
Astropy Coordination Meeting, input was gathered from participants on what governance
structures existed in the associated Open Source Software communities, and what
would fit well with the needs of \astropy. This led to a ``retreat'' planned
for March 2020, but due to the COVID-19 pandemic, this became a series of
virtual meetings of the ``Astropy Governance Working Group.'' This group
drafted the APE 0 document \citep{ape0}, which was eventually ratified and
implemented by the ``Astropy Governance Implementation Working Group'' in Fall
2021. While the process emphasized flexibility and adaptability,
it is the expectation that \astropy's governance will operate in
this framework for at least the medium-term future.

The APE 0 \citep{ape0} document lays out the principles of this governance
structure, so we refer the reader to that document for a more thorough
description, and here we only highlight some key elements. The APE 0-based
governance formalizes the building blocks that are already de facto true or
previously discussed.
With this in mind, it highlights the developer and user community of the
\astropy Project as the ultimate sources of authority, as well as the core
principle of ``do-ocracy'': those who do work for the Project (coding,
training, or other harder-to-quantify contributions) gain more influence on the
outputs of the Project by virtue of their effort. However, APE 0 adds the
concept of ``voting members'': a self-governed part of that community who are
entrusted to elect the Coordination Committee (CoCo). While the CoCo
predates APE 0, membership, rights, and responsibilities are now formally outlined.
This role is mainly to facilitate consensus and make decision
when other mechanisms have failed, and includes powers that either
require central authority or secrets (e.g., passwords). However, APE 0 also
charges the Committee to devolve responsibilities and seek community input on
these items as often as possible.

The first CoCo election took place in Fall 2021, electing a mix of prior and new members.
New roles and voting members were also established. This
suggests the process is already working to serve the long-term interests of the
committee to both spread the coordination effort, and to ensure it is not
dominated by the same people for as long as the Project continues. While other,
more fine-grained governance improvements are planned for the future, it is
clear the foundation is now in place.

\subsection{Core package contributor base} \label{sec:project-contributors}

% \secauthor{Adrian Price-Whelan}

% TODO: move this subsection to "updates to core package" section above?

\begin{figure}[t!]
    \begin{centering}
      \includegraphics[width=0.9\textwidth]{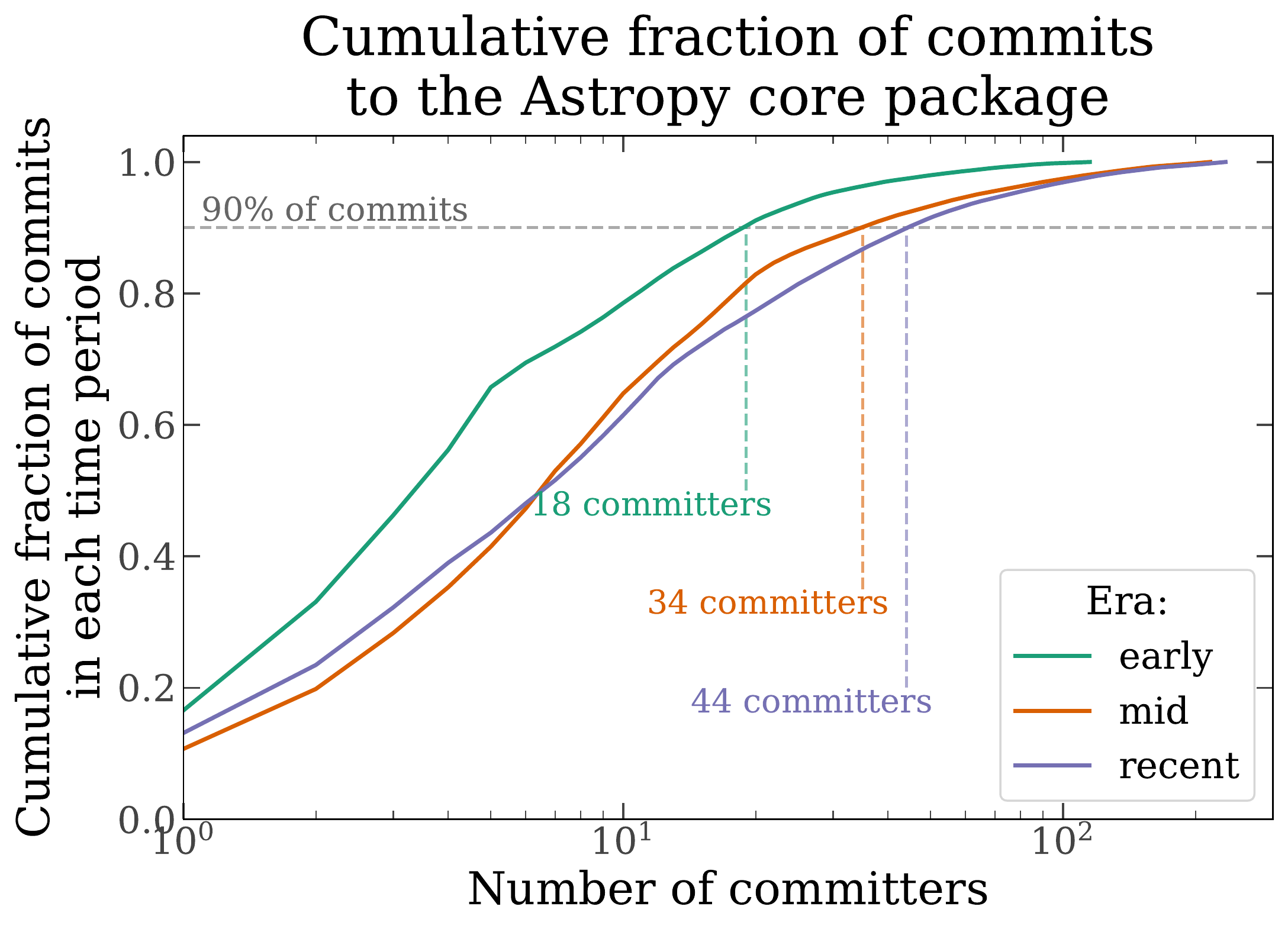}
        \caption{
            The cumulative fraction of Git commits to the \astropypkg core
            package as a function of number of contributors (committers) broken
            into three equal-length time periods between the start of \astropypkg
            and the feature freeze date of v5.0.
            The time eras early, mid, and recent correspond to the approximate
            year ranges 2011--2015, 2015--2018, 2018--2021 (defined more
            precisely in Section~\ref{sec:project-contributors}).
            In all eras, the majority of commits to the \astropypkg core package
            have come from a small subset of the total number of contributors.
            For example, as annotated in the figure, 90\% of the Git commits
            have come from 18, 34, and 44 committers in the early, mid, and
            recent eras, respectively.
        }
        \label{fig:contributor-summary:Ncommitters}
    \end{centering}
\end{figure}

As described in the introduction above, the \astropy Project is broader than the
core \astropypkg package as it also represents a number of core infrastructure,
coordinated, and affiliated packages, educational initiatives, and user communities.
In total, well over 1500 people have contributed in some way to the Project,
whether by contributing code or documentation, reporting issues
on \github, providing infrastructure support, writing tutorials, or helping to
coordinate \astropy workshops.
The vast majority of these contributors and participants have interacted as
volunteers choosing to engage with the open-source and open-development software
community.
However, while the total number of contributors is large, the number of
contributions and bulk of the maintenance of the \astropy Project components has
primarily been sustained by a small number of heavily-engaged or
explicitly-employed contributors \citep[e.g.,][]{astropy:2018}, as is the
case in many large, open-source software communities
\citep[e.g.,][]{numpy:nature}.

As an example of this within the \astropy Project, here we focus on
contributions made to the \astropypkg core package.
Figure~\ref{fig:contributor-summary:Ncommitters} shows the number of
contributors who have contributed a given fraction of the total number of Git
commits, separated into three separate time spans (``eras'') in the history of
the Project.\footnote{Note, however, that measuring these statistics in terms of
Git commits is not a perfect way of encapsulating ``effort,'' but we use this
metric for its simplicity and clear definition.}
The annotations show that, within the early (July 2011--January 2015), mid
(January 2015--June 2018), and recent (June 2018--November 2021) eras, 90\% of
the commits to the \astropypkg core package have come from 18, 34, and 44
contributors.
While the trend is encouraging --- a larger pool of contributors are
contributing more, fractionally --- this emphasizes that, still, the vast
majority of the work in \astropypkg is done by a small subset ($<20\%$) of the
115, 216, and 234 contributors, respectively, within each era.

\begin{figure}[t!]
    \begin{centering}
      \includegraphics[width=0.75\textwidth]{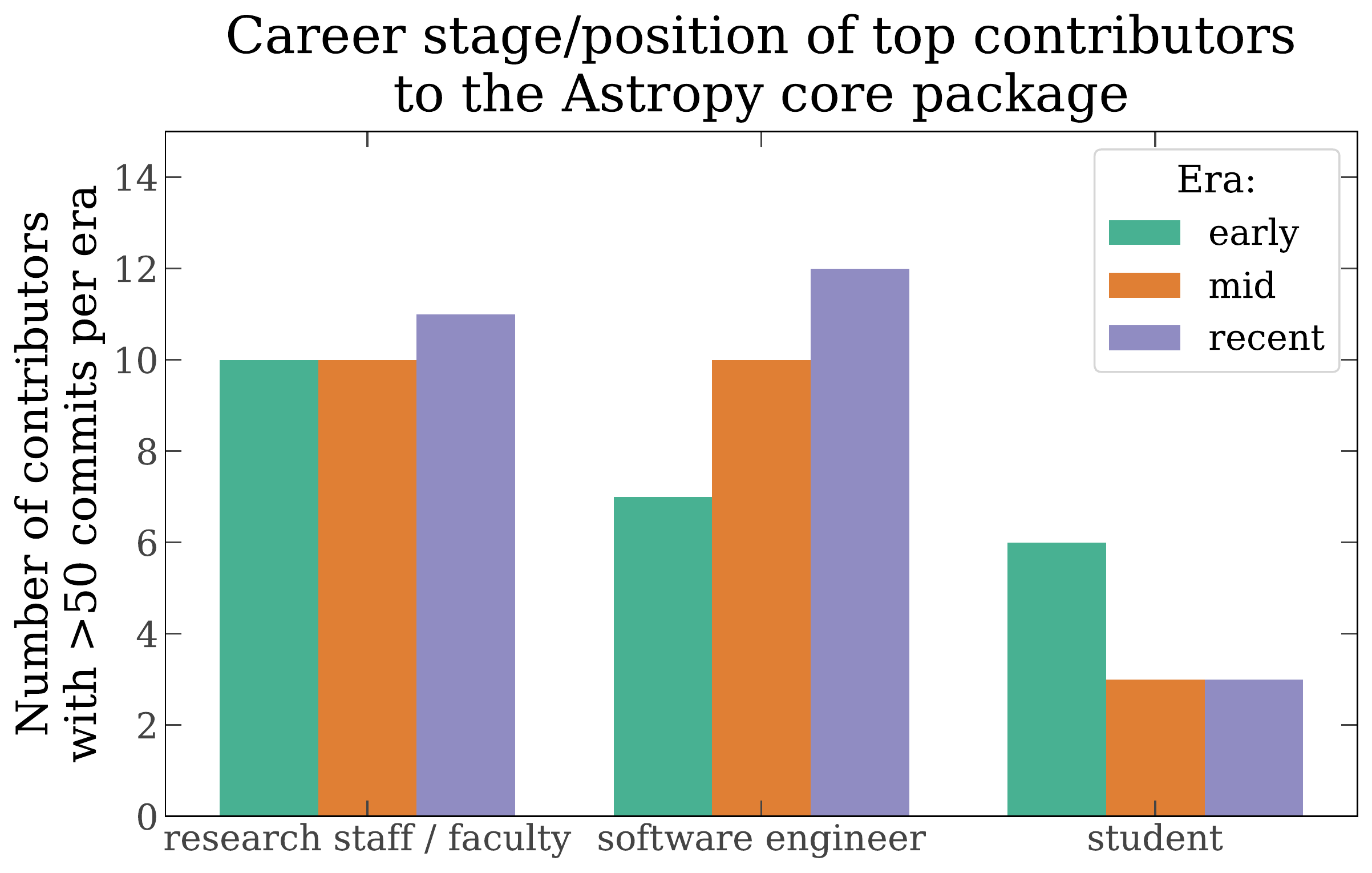}
        \caption{
            The career stages and/or positions of the top contributors within
            each time era (defined as having contributed $>50$ commits to the
            \astropypkg core package in each era).
            The time eras early, mid, and recent correspond to the approximate
            year ranges 2011--2015, 2015--2018, 2018--2021 (defined more
            precisely in Section~\ref{sec:project-contributors}).
        }
        \label{fig:contributor-summary:top}
    \end{centering}
\end{figure}

Since its inception, the most active contributors to the \astropypkg package
have spanned a range of career stages and paths.
For example, Figure~\ref{fig:contributor-summary:top} shows a breakdown of the
career stage/path at the midpoint of the three time eras defined in the previous
paragraph for all contributors with more than 50 commits to the \astropypkg
package within each era.
Over the $\sim$6 years between the ``early'' and ``recent'' eras, many
contributors moved between these categories (i.e., have graduated as students to
become research staff/faculty or software engineers), but also a number of
early contributors moved on to other positions and stopped contributing to the
core \astropypkg package.
Given this, it is encouraging that the numbers in each career stage/path have
stayed relatively constant over the history of \astropy (with some showing
growth), as new, active contributors have generally joined and stepped in to
fill top-contributor roles within the Project.
On the other hand, the total number of top contributors has not grown
substantially, reflecting one of our principal challenges: sustaining a pipeline
from users to contributors to maintainers of the package.
One other thing to note in this context is that in all eras, 50--60\% of the
software engineers who contributed to \astropypkg were (or are) employed by
STScI, whose leadership has consistently provided resources (financial and
personnel) towards sustaining \astropy.

\begin{figure}[t!]
    \begin{centering}
      \includegraphics[width=0.75\textwidth]{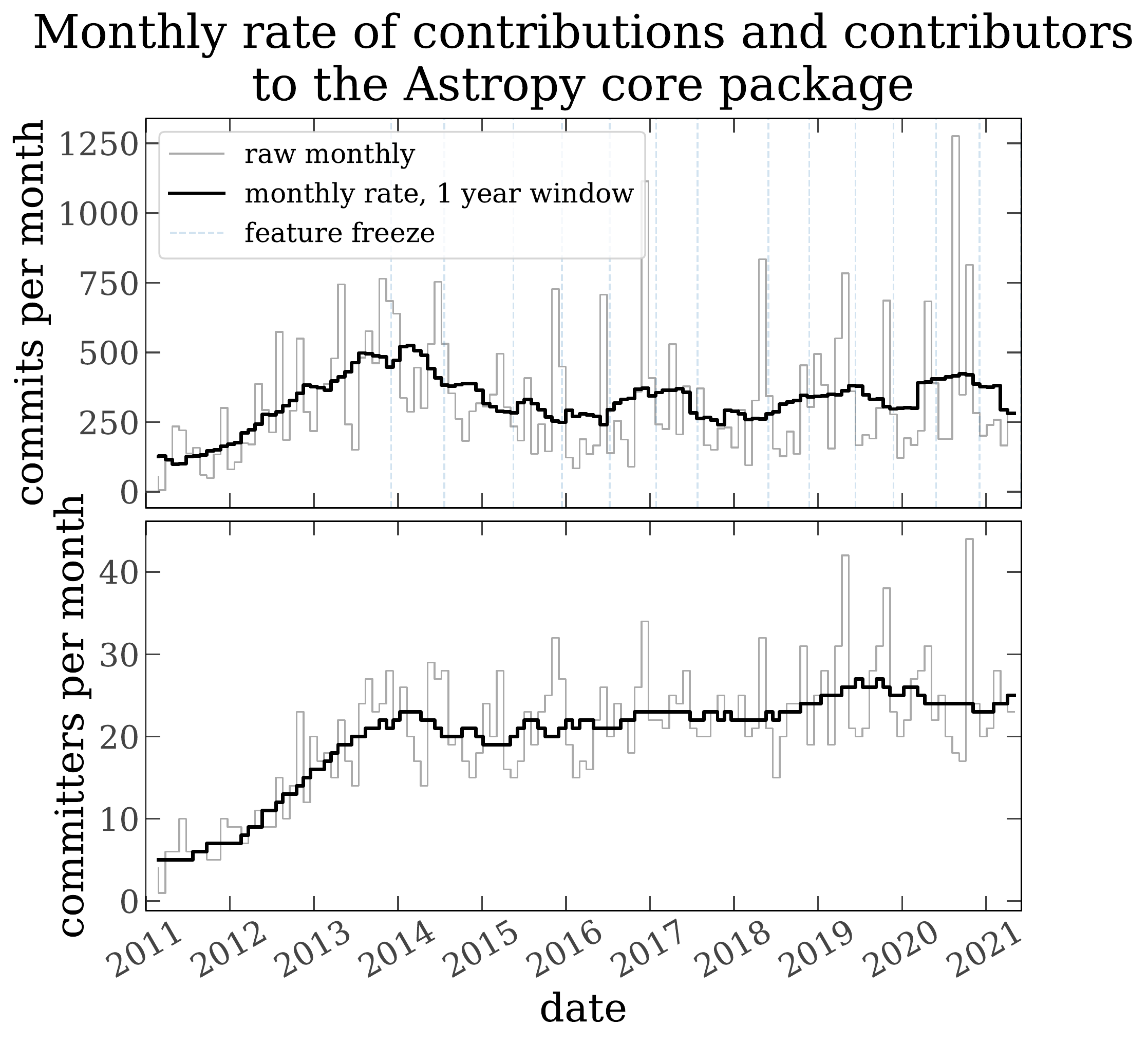}
        \caption{
            \textbf{Top panel:} The number of Git commits per month (defined as
            consecutive 30 day windows) merged into the \astropypkg core
            package over the whole history of the Project (up to end of 2021,
            v5.0).
            The dark line shows the mean rate using a 1 year rolling window, but
            the under-plotted gray line shows the raw monthly rate.
            The raw number clearly peaks just before feature freezes (vertical
            lines) as a flurry of activity leading up to each release, but in
            recent years the mean rate has remained steady at about $\sim$300
            commits per month.
            \textbf{Bottom panel:} Similar to above, but now showing the number
            of unique Git committers who contribute to the \astropypkg core
            package each month (again defined as consecutive 30 day windows).
            While this also shows some variation depending on proximity to
            feature freezes, in recent years the mean rate of unique
            contributors has remained consistent at about $\sim$22 committers
            per month.
        }
        \label{fig:contributor-summary:per-month}
    \end{centering}
\end{figure}

A final metric we consider as a diagnostic of the contributor base and
contributions to the \astropypkg core package is the rate of contributions and
contributor participation, defined again using Git commits to quantify
contributions.
Figure~\ref{fig:contributor-summary:per-month} shows the monthly rates of
commits and unique committers to the \astropypkg core package over the history
of the package.
After an initial ramp-up in the commit rate and number of committers (lasting
roughly from 2011 to 2014), the commit rate and number of unique committers have
both remained fairly constant.

% Transition of needs: From new features, to community development, sustainability, ...
% - pipeline between User, Participant, Code contributor, Maintainer, Coordinator.

\subsection{Inclusion, Diversity, and Equity Programs} \label{sec:project-ide}

% \secauthor{Lía Corrales}

With support from the Moore Foundation, the \astropy Project was
allocated funding to support mentoring programs. In 2020 a call was
made to submit proposals for Inclusion, Diversity, and
Empowerment (IDE) initiatives for project-wide consideration on
\github. This process was deemed the most ``open'' because it allowed
for community-wide feedback to focus and improve proposal
initiatives. Two highlighted programs that were selected and implemented via
this process are described below.

% \textbf{Outreachy:} \todo{No author assigned}{Describe the outreachy project}

\textbf{Women of Color Code (WoCCode)} is a peer-mentoring network for
coders from traditionally marginalized groups, most notably women of
color. Participants were invited to the WoCCode Slack space and
encouraged to attend monthly webinars to share skills in the context
of open-source software libraries. Every other month, a guest speaker
was invited to talk about their career path and share a skill. Program
participants were solicited in the fall of 2020, yielding 73
applications from 17 countries (48\% from the United States, 37\% from
Africa, and 15\% from remaining continents). We selected participants
who we identified as having a high potential for contributing to
open-source projects: intermediate to advanced programming skills with a
vocal interest in contributing. Of the thirty applicants invited to
the program, nineteen joined the community on Slack. Participants were
organized into cohorts based on interest and each cohort was assigned
one of three mentors that were also selected via open
application. Mentors acted as a general resource to participants, gave
one webinar, and organized a hack day. WoCCode also supported
registration of two participants to attend the American Astronomical
Society virtual summer 2021 meeting.

Participants rated the impact of the program as very high. In the
final webinar, participants reported a change in their perception of
coding in general, for example, accepting coding as something they can
do for fun. Participants reported generally feeling comfortable asking
questions and interacting with a community where ``[e]veryone's
thoughts are welcomed, no one is made to feel less important'' and
where one can enjoy ``warm interactions with likeminded people.''
WoCCode is continuing into 2022 by broadening participation in the
Slack space and publicly advertising the guest webinar events.

In addition to the supported mentoring initiatives, the \astropy
Project as a whole has taken steps to examine representation of
marginalized groups within the Project and search for avenues of
improvement.

% Much of this text was adapted from Robel Geda's report from NSBP and
% SACNAS. Please include him on the author list.
\textbf{\astropy representation at national diversity conferences:}
Several members of the \astropy community attended virtual conferences of the
National Society of Black Physicists (NSBP) and SACNAS (focusing on Hispanic and
Native American scientists across all STEM fields). \astropy representatives
noted that one underlying topic came up multiple times: a major barrier for
persons who come from underrepresented communities is the lack of pedagogical
resources and expertise. For instance, software engineering skills and best
practices are not part of the standard
Physics and Astrophysics curriculum. Additionally, the larger astronomical
community is still in the process of transitioning towards \python-based tools,
hindering students not currently under advisement by someone with
extensive \python expertise from getting involved with the open-source \python
community. Projects such as Learn Astropy (see Section~\ref{sec:learn}) could
have a profound impact by empowering underrepresented groups because it provides
a free, searchable, and accessible introduction to \python tools for
astrophysical research. Representatives also noted that offering training and
teaching materials to PIs of Research Experience for Undergraduate (REU)
programs throughout the United States could be helpful for encouraging advisors
to teach students how to use Astropy, create their own libraries, and use
version control on open-source platforms like \github. Such materials could also
be offered as a workshop at national diversity conferences themselves, as SACNAS
solicits special session proposals each year.

% \textbf{NumFOCUS ``Contributor Diversification \& Retention'' (CDR)
%   initiative:} \todo{No author assigned}{Describe more about this.}

\subsection{Project funding} \label{sec:project-funding}

% \secauthor{Aarya Patil}

% Summarize funding sources (Moore, NASA) and amounts and what this has been
% used for.

With the goal of establishing a sustainable financial model for the \astropy
Project, the CoCo, under the new governance charter (see
Section~\ref{sec:project-governance}), established a standing Finance Committee.
This committee oversees the planning and allocation of finances on behalf
of the Project. The \astropy Project accepts funds from institutions as well as
individuals through NumFOCUS.\footnote{NumFOCUS is a 501(c)(3) non-profit
organization in the United States of America that can collect donations,
administer grants, and perform financial management for projects in open source
scientific computing.}
Previously, no direct financial support was available for project development,
and NumFOCUS covered most of the incurred operational costs. However, transitioning
to long-term financial stability meant having an influx of funds in the form of
institutional support, and hence the \astropy Project applied for and successfully
acquired four grants over the last several years. These grants are administered by NumFOCUS on behalf of the
Project, providing a ``neutral'' space not tied to any particular astronomy
institution. The two major grants are briefly summarized below.

\begin{description}
    \item[Moore] The Gordon and Betty Moore Foundation awarded the \astropy
    Project a $\sim$\$900k (US) grant in 2019 for the proposal ``Sustaining and
    Growing the Astropy Project.'' Over a three-year term, this grant supported
    sustenance of the Project by providing funds for (1) the development and
    maintenance of \astropypkg and its infrastructure, provided to existing
    project members and targeted hires, (2) the transition of long-term users to
    contributors and maintainers through mentorship efforts, (3) the
    formalization of a governance structure (see
    Section~\ref{sec:project-governance}), and (4) the improvement of equity,
    diversity, and inclusion efforts in the Project (see
    Section~\ref{sec:project-ide}). Some of this money supported travel for
    meetings, conferences, and workshops. This generous grant helped reduce
    reliance on volunteer-driven maintenance of the openly-developed package and
    paved the way to acquire funding from US federal agencies, representing a
    major milestone for the success of the Project.

    \item[NASA] A successful proposal to the NASA ROSES-2020
    call\footnote{\url{http://solicitation.nasaprs.com/ROSES2020}}, section E.7
    (Support for Open Source Tools, Frameworks, and Libraries), granted the
    \astropy Project $\sim$\$600k (US) to support its work. Funded work for a
    term of three years includes (1) project-wide infrastructure maintenance and
    improvements, (2) targeted work on areas of the \astropy core and
    coordinated packages, (3) enhancements to the Learn
    Astropy ecosystem and educational materials (see Section~\ref{sec:learn}),
    and (4) support for the \astropy affiliated packages. These goals
    support the sustenance of the Project by encouraging further
    development of the package ecosystem and community engagement.
\end{description}

In addition to the two grants described above, the Project benefited from
a Gemini Observatory contract under its Science User Support Department,
which awarded funds for development work that supports both the Astropy and
DRAGONS\footnote{Data Reduction for Astronomy from Gemini Observatory North and
South (\url{https://github.com/GeminiDRSoftware/DRAGONS}).} projects. The
Project also received financial support from the Dunlap
Seed Funding program at the University of Toronto to develop educational
resources for scientific software packaging within the Learn Astropy
framework. Moreover, the Project indirectly benefited from the NRAO ALMA Development
Study program, which supported the grant ``Linking CASA to the astropy
ecosystem''\footnote{\url{https://science.nrao.edu/facilities/alma/science_sustainability/Spectral_Cube_and_Radio_Beam_Ginsburg.pdf}}.
All of these successful funding proposals show that the \astropy Project
is vital for the astronomical infrastructure and community at large, and
strengthen its promising and sustainable future.

The Project follows a set of principles to fairly and effectively
distribute the above described funds. By drawing inspiration from
Numpy's NEP 48\footnote{https://numpy.org/neps/nep-0048-spending-project-funds.html},
the APE 19 \citep{ape19} document lays out these principles. We refer the
reader to that document for more details.

\section{Supporting the Ecosystem of Astronomical Python Software}
\label{sec:ecosystem}

\subsection{Community-oriented infrastructure}

% \secauthor{Nicholas Earl}

The Astropy Project supports the broader ecosystem by providing pre-configured
infrastructure packages that the community can use to support and maintain their
own software package infrastructure. These include tools and extensions that
enable generating documentation and setting up automated testing, as well
as providing package scaffolding for new projects.

Sphinx is a common and useful tool for generating documentation for \python
packages. The Astropy Project maintains a default Sphinx configuration along
with Astropy-specific extensions that can be easily added to community
projects via the
\href{https://github.com/astropy/sphinx-astropy}{\texttt{sphinx-astropy}}
meta-package. This tool provides a pre-configured Sphinx setup compatible with
Astropy projects, which includes several extensions useful for generating API
documentation
(\href{https://github.com/astropy/sphinx-automodapi}{\texttt{sphinx-automodapi}}),
allowing for \package{numpy} docstring parsing
(\href{https://github.com/numpy/numpydoc}{\texttt{numpydoc}}), embedded image
handling
(\href{https://github.com/sphinx-gallery/sphinx-gallery}{\texttt{sphinx-gallery}};
\href{https://github.com/python-pillow}{\texttt{pillow}}), advanced
documentation testing support
(\href{https://github.com/astropy/pytest-doctestplus}{\texttt{pytest-doctestplus}}),
and providing a custom documentation theme ideal for analysis packages
(\href{https://github.com/astropy/sphinx-astropy}{\texttt{astropy-sphinx-theme}}).

Community package testing infrastructure is supported through the
\href{https://github.com/astropy/pytest-astropy}{\texttt{pytest-astropy}}
meta-package, providing a unified testing framework with useful extensions
compatible with both Astropy- and non-Astropy-affiliated community packages.
This meta-package pulls in several
\href{https://github.com/pytest-dev/pytest}{\texttt{pytest}} plugins to help
with custom test headers
(\href{https://github.com/astropy/pytest-astropy-header}{\texttt{pytest-astropy-header}}),
accessing remotely-hosted data files in tests
(\href{https://github.com/astropy/pytest-remotedata}{\texttt{pytest-remotedata}}),
interoperability with documentation
(\href{https://github.com/astropy/pytest-doctestplus}{\texttt{pytest-doctestplus}}),
dangling file handle checking
(\href{https://github.com/astropy/pytest-openfiles}{\texttt{pytest-openfiles}}),
data array comparison support in tests
(\href{https://github.com/astropy/pytest-arraydiff}{\texttt{pytest-arraydiff}}),
subpackage command-line testing support
(\href{https://github.com/astropy/pytest-filter-subpackage}{\texttt{pytest-filter-subpackage}}),
improved mock object testing
(\href{https://github.com/pytest-dev/pytest-mock}{\texttt{pytest-mock}}), test
coverage reports and measurements
(\href{https://github.com/pytest-dev/pytest-cov}{\texttt{pytest-cov}}), and
configuring packages for property-based testing
(\href{https://github.com/HypothesisWorks/hypothesis}{\texttt{hypothesis}}).

The Astropy Package Template helps facilitate the setup and creation of new
\python packages leveraging the Astropy ecosystem. This tool utilizes the
\href{https://github.com/cookiecutter/cookiecutter}{Cookiecutter} project to
walk users through the process of creating new packages complete with
documentation and testing support. Additionally, the package template
generation process includes the ability to setup interoperability with
\github, allowing for easy repository access from documentation, as well as an
example \github Actions workflow to demonstrate the use of \github's
continuous integration tooling.

\subsection{Astropy affiliated packages}

% \secauthor{Matt Craig, Brett Morris}

% Highlight a few new affiliated packages and major updates to existing ones.
% Include a big table of all affiliated packages and references (as in v2.0
% paper).

As defined in \paperii, Astropy affiliated packages are astronomy-related
\python packages that provide functionality that builds upon and extends the
\astropypkg core package that have requested to be included as part of the
Astropy Project community. These packages support the goals and vision of
Astropy of improving code re-use, interoperability, and embracing good coding
practices such as testing and thorough documentation, but are not solely
developed and maintained by members of the Astropy Project team.
However, a subset of the affiliated packages --- the ``Astropy Coordinated
Packages'' --- are maintained by the Project.
All Astropy-coordinated and other affiliated packages are listed with detailed
information on the \href{{https://www.astropy.org/affiliated}}{\astropy
website}; a table summarizing key aspects of the affiliated packages is included
in the Appendix below (Table~\ref{tab:affiliated-registry}).

Since \paperii, there has been an expansion of affiliated
packages for gravitational astrophysics, including:
\texttt{PyCBC} \citep{PyCBC} for exploring gravitational wave signals, \texttt{lenstronomy} \citep{lenstronomy:1} for
modeling strong gravitational lenses, \texttt{ligo.skymap} for visualizing
gravitational wave probability maps, and \texttt{EinsteinPy} \citep{EinsteinPy} for general
relativity and gravitational astronomy. There have also been several packages
added to the ecosystem related to HEALPix: \texttt{astropy-healpix}
for a BSD-licensed HEALPix implementation, and \texttt{mocpy} for Multi-Order
Coverage maps. \texttt{astroalign}  \citep{astroalign} has been introduced for astrometric registration,
and \texttt{python-cpl} products \citep{pythoncpl} has been added for ESO pipelines and VLT data.
For ground-based astronomy, \texttt{baseband} \citep{baseband} has added IO capabilities for
VLBI, and \texttt{SpectraPy} \citep{SpectraPy} brings slit spectroscopy to the \astropy ecosystem.
Other new affiliated packages include \texttt{agnpy} \citep{agnpy} for AGN jets,
\texttt{statmorph}  \citep{statmorph}for fitting galactic morphological diagnostics,
\texttt{dust\_extinction} for modeling interstellar dust extinction,
\texttt{feets} for extracting features from time series data, and
\texttt{corral} \citep{corral} for managing data intensive parallel pipelines.
\texttt{casaformats-io} provides infrastructure for reading data stored in CASA table formats.
\texttt{saba} gives an interface to the Sherpa \citep{sherpa} fitting routines,
\texttt{BayesicFitting} \citep{BayesicFitting} provides an interface for generic Bayesian inference,
and \texttt{sbpy} \citep{sbpy} enables calculations for asteroid and
cometary astrophysics. Finally, \texttt{synphot} \citep{synphot} provides an interface for
synthetic photometry.
% \citep{Mommert2019}

Several of the coordinated and affiliated packages described in
\citet{astropy:2018} have had substantial improvements. \texttt{astroquery}
\citep{astroquery} has added access to roughly a dozen new missions and data
services, including the JWST archive, and the
project has switched to a continuous release model: every time a change is
committed to the main development branch it is published on the Python Package
Index (PyPI) and available for installation. (However, formal releases are still
done a few times per year.) \texttt{photutils} \citep{photutils} released its
first stable version, indicating that the API will change less frequently, and
there have been several significant performance improvements. \texttt{ccdproc}
\citep{ccdproc}  released a new major version, bringing better performance to
some image combination operations. The \texttt{regions} package
\citep{regions}, for manipulating ds9-, CASA-, or FITS-style region definitions \citep{ds9},
added new ways to manipulate regions, introduced new region types, and obtained
expanded visualization and interactive region editing capabilities.
\texttt{reproject}'s \citep{reproject} major new feature is a function to align
and co-add images to create a mosaic; better support for parallelization was
also added. \texttt{specutils}, the package that defines containers for 1D and
2D spectra \citep{specutils},  had its first stable release and the addition
of classes to read JWST data. \texttt{stingray} \citep{stingray:2} has had a major performance
overhaul. The package \texttt{gammapy} \citep{gammapy} has had major performance
improvements and has unified its API in preparation for its first stable
release.

\subsection{Connections with data archives}
\label{sec:astroquery}

% \secauthor{Adam Ginsburg}

\texttt{astroquery} \citep{Ginsburg2019} is the Astropy-coordinated package for
interacting with online archives of astronomical and related data. It contains
over 50 modules for querying astronomical databases, large and small. In
particular, since \paperii, significant contributions to \texttt{astroquery}
have come from several of the major archives, including the European Space
Agency (ESA), the Mikulski Archive for Space Telescopes (MAST) at the Space
Telescope Science Institute (STScI), the Infrared Science Archive at the NASA
Infrared Processing and Analysis Center (IRSA at NASA IPAC), the Canadian
Astronomical Data Center (CADC), and the Atacama Large Millimeter/Submillimeter
Array archive (ALMA). These contributions represent a formal acknowledgement of
the utility of a centralized tool suite for archive interaction from \python.
The widespread usage of \texttt{astroquery} is apparent from the range of
keywords represented in the journal articles that cite the \texttt{astroquery}
paper \citep{Ginsburg2019}; everything from asteroids to galaxies is
represented. Additionally \texttt{astroquery} functionality has been built into
several special-purpose \python packages, such as \texttt{LightKurve}
\citep{LightKurve} and SORA \citep{SORA}. More than 4,000 repositories on
\github make use of \texttt{astroquery} in some way, and \texttt{astroquery}
features in a large number of tutorials on various facets of astronomical
analysis.

Many of the existing and newly-contributed tools rely on Virtual Observatory
(VO) tools. These use the underlying package \texttt{pyvo}, which has also
recently become an Astropy-coordinated package. While many or all of the
functions provided in \texttt{astroquery} can be achieved through direct use of
VO tools implemented in \texttt{pyvo}, the \texttt{astroquery} interfaces more
closely resemble the web interfaces that are more familiar to most users.

\subsection{Connections with Observatories and Missions}

In this section, we highlight a few efforts in observatory- or mission-driven
development that have contributed to Astropy, and vice versa.

\subsubsection{JWST}
% \secauthor{Larry Bradley}

JWST is a 6.5-meter space-based
infrared telescope that will provide unprecedented resolution and
sensitivity from 0.6--28 microns. JWST will enable a broad range
of scientific investigations from exoplanets and their atmospheres
to the formation of galaxies in the very early universe. Its four
key scientific goals are to study the first light from stars and
galaxies, the assembly and evolution of galaxies, the birth of stars and
protoplanetary systems, and planetary systems and the origins of life.

The telescope launched on an Ariane 5 rocket on 2021 December 25 from
Kourou, French Guiana. After a series of successful deployments,
including the sunshield and primary and secondary mirrors, JWST reached
its orbit around the L2 Lagrange point on 2022 January 24. Commissioning
of the telescope optics and science instruments will occur from January
until the end of June 2022, when science operations are scheduled to
begin.

Software developers at the Space Telescope Science Institute (STScI),
the operations center of JWST, have been developing \python-based tools
for JWST since 2010 (starting with the JWST Calibration Reference Data
System) and have provided major contributions to \astropy from its
inception. The JWST instrument calibration pipelines, exposure-time
calculators, and data analysis tools are all written in \python and
depend on the \astropypkg core package and some coordinated and
affiliated packages. For the \astropypkg core package, the JWST mission
has provided extensive contributions to the \astropysubpkg{modeling},
\astropysubpkg{units}, \astropysubpkg{coordinates}, \astropysubpkg{wcs},
\astropysubpkg{io.fits}, \astropysubpkg{io.votable},
\astropysubpkg{stats}, \astropysubpkg{visualization}, and
\astropysubpkg{convolution} subpackages as well as to the general
package infrastructure and maintenance.

Likewise, JWST developers have provided significant contributions
to the several affiliated packages. For example, development of the
\package{photutils} coordinated package for source detection and
photometry has largely been led by JWST contributions. JWST developers
have also made significant contributions to the \package{specutils}
coordinated package, which is used for analyzing spectroscopic data,
and the \package{regions} coordinated package, which is used to
handle geometric regions. The \package{gwcs} affiliated package for
generalized world coordinate systems was created specifically to handle
the complex world coordinate systems needed for JWST spectroscopic
data. The \package{synphot} affiliated package for synthetic photometry
was created at STScI and is a dependency of the JWST exposure-time
calculators. Furthermore, the \package{ASDF} (Advanced Scientific Data
Format) package \citep{ASDF}, a next-generation interchange format for
scientific data, was initially developed at STScI to serialize JWST WCS
objects along with \astropypkg models, units, and coordinates. Over
its mission lifetime, JWST will continue its support in developing and
maintaining these critically dependent packages.

% Gemini
% \secunfilled

\subsubsection{Cherenkov Telescope Array}
% \secauthor{Axel Donath, Maximilian Nöthe}

The Cherenkov Telescope (CTA) will be the next generation very-high-energy
gamma-ray observatory.
CTA will improve over the current generation of imaging atmospheric Cherenkov telescopes (IACTs)
by a factor of five to ten in sensitivity and will be able to observe the whole sky from a combination of two sites:
a northern site in La Palma, Spain, and a southern one in Paranal, Chile.
CTA will be able to observe gamma rays in a broad energy range from around $20\,\mathrm{GeV}$ to over $300\,\mathrm{TeV}$
using three different types of telescopes. In total over 100 telescopes are planned at the two sites.
CTA will also be the first open gamma-ray observatory.

The data analysis pipeline is developed as open-source software and essentially split in two domains:
\begin{enumerate}
  \item In the low-level analysis, the properties of the recorded air-shower events
    have to be estimated from the raw data.
    The raw data consists of very short ($\sim40$--$100\,\mathrm{ns}$) videos recorded with the fast and
    sensitive cameras of the telescopes.
    This includes the energy, particle type and direction of origin of the particle that induced the air shower
    and the time the shower was recorded.
  \item In the higher-level analysis, these reconstructed event lists are used together with some
    characterization of the instrument response to perform the actual scientific analysis.
    This software will be delivered as CTA science tools to the future users of the Observatory.
\end{enumerate}

A prototype for the low-level analysis is \texttt{ctapipe} \citep{ctapipe},
a \python package developed to perform all the necessary tasks to perform data processing, from the raw data
of Cherenkov telescopes to the reconstructed event lists.
The high-level analysis (or CTA science tools) will be based on the Astropy affiliated package
\package{gammapy}~\citep{gammapy}.
Both \package{gammapy} and \texttt{ctapipe} make heavy use of the \astropy core package, mainly for units, times, coordinate transformations,
tables, and FITS I/O.
As CTA will record gamma-ray events with a rate of up to $10\,000$ events per second,
it needs to perform a large number of coordinate transformations.
To enable this, CTA member M.~Nöthe contributed a major performance improvement
for large numbers of coordinates with different observation times,
based on earlier work by B.~Winkel.
Together with \package{gammapy} maintainer A.~Donath and former maintainer C.~Deil,
a total of 107 merged pull requests were contributed to astropy.

\subsubsection{Data Central and the Anglo-Australian Telescope Archive}
% \secauthor{James Tocknell}
Data Central is one of the five Australian Virtual Observatory\footnote{Known as
  the All-Sky Virtual Observatory, or ASVO.} nodes, providing both Australian and
international astronomers access to and services for surveys of the southern
sky. Whilst primarily hosting optical data, including the Anglo-Australian
Telescope (AAT) Archive\footnote{The AAT Archive is currently jointly managed by
  Data Central, Astronomy Australia Limited (AAL), and National Computational
  Infrastructure (NCI).}, Data Central also assists theoretical and radio
astronomers with hosting and managing their data.

\python is widely used by Data Central, and \astropy and its wider ecosystem
play a vital role in the initial ingestion, reduction, access and visualization
of the hosted data and services. For example, the AAT Archive hosts data from as early as
1974, so \astropycoordinates, \astropytime and \astropyfits are used to ingest
data flowing from the AAT to ensure that a consistent time and coordinate system
is provided to users for simpler querying.

As many of the surveys hosted by Data Central are spectroscopic surveys from the
AAT, \package{specutils} is heavily used to standardize the variety of encodings
of spectroscopic data into a form easily readable and visualizable by many
standard tools (such as \package{ds9} or \package{splat}). Data Central has
upstreamed into \package{specutils} loaders for all hosted spectra, including
\package{2dfdr}-reduced spectra from the AAT Archive, and strongly encourages other
archives to provide loaders for their data to further the reuse of existing archival
data.

% Rubin Observatory
% \secunfilled

% LIGO/Virgo/KAGRA
% \secauthor{Leo Singer}

\subsection{Learn Astropy} \label{sec:learn}

% \secauthor{Lía Corrales, David Shupe, Kelle Cruz + Learn team}

% \subsubsection{Current status and scope}

\textit{Learn Astropy} is an umbrella term that acknowledges the broad
educational efforts made by the Astropy Project, which are led by the Learn
Astropy Team.
The efforts focus on developing online content and workshops covering
astronomy-specific coding tasks in \python.
As introduced in \citet{astropy:2018}, there are four different types of
Learn Astropy content: \textit{tutorials}, consisting of Jupyter Notebook
lessons that are published in HTML format online; \textit{guides}, which are a
series of lessons providing a foundational resource for performing certain
type of astronomical analyses; \textit{examples}, which are snippets of code
that showcase a short task that can be performed with \astropypkg or an affiliated
package; and \textit{documentation}, which contains more detailed information
about the code base and user interface.
This categorization drives content development, infrastructure choices, and the
appearance of the \href{https://learn.astropy.org}{Learn Astropy website}.
The Learn Team meets weekly to work on creating, expanding, and improving these
educational resources.

The Learn Team recently re-launched the main website and search interface for
Learn Astropy in 2021 with a new infrastructure platform, built around full-text
search and interactive filtering functionality, with the goal of making content
more easily searchable and discoverable as the Learn Astropy content catalog
expands.
This work has been supported in part by a grant from the Dunlap Institute.
We have adopted Algolia, a search-as-a-service cloud platform, to store the
full-text and metadata records of Learn Astropy's content.
The new Learn Astropy website is a JavaScript (Gatsby/React) application that
uses the Algolia service to power its search and filtering user interface.
Our \python-based application,
\href{https://github.com/astropy/learn-astropy-librarian/}{Learn Astropy
Librarian}, populates data into the Algolia service. We tuned the Librarian
around specific content formats (such as Jupyter Notebook-based tutorial pages
and Jupyter Book-based guides) to more accurately index content and
heuristically extract metadata.
A consequence of the new platform is that we now maintain and compile content
separately from the website application itself, enabling new content types.
Tutorials, which are written as Jupyter Notebooks, are now compiled into their
own Learn Astropy website using
\href{https://github.com/astropy/nbcollection}{\texttt{nbcollection}}.
Guides, which utilize the \href{https://jupyterbook.org/}{Jupyter Book} build
infrastructure, are also deployed as separate websites using \github pages.
This architecture opens future possibilities of indexing third-party content,
hosted elsewhere, such as on institutional websites.

We currently host 19 tutorials written as interactive Jupyter Notebooks and
rendered into static HTML pages with the infrastructure described above.
The tutorials span a range of astronomical topics, from general tasks like
reading or creating FITS files with different content or working with
astronomical coordinate systems, to more specific exercises like analyzing
spectroscopic data from the UVES instrument.
Now that the backend infrastructure is stable, we are interested in collecting
new content to serve and share.
Ideas for contributions, new tutorials, or suggestions for new tutorial series
to be made searchable through the Learn Astropy interface can be raised by
creating a \github issue in the
\href{https://github.com/astropy/astropy-tutorials}{Astropy Tutorials GitHub
repository}.
We currently host one Guide --- a longer-form walkthrough of a more complex
concept or topic --- that is focused on
\href{http://www.astropy.org/ccd-reduction-and-photometry-guide}{CCD image
reduction and photometry}, but are also actively seeking new material that would
be suitable for new guides.

Beyond developing and serving educational content, the \astropy Project has been
conducting workshops at winter meetings of the American Astronomical Society
since AAS 225 in January 2015.
Up to the start of the coronavirus pandemic, these were full-day in-person
workshops with as many as 90 participants and a dozen facilitators from the
Project.
During the pandemic, these workshops were moved to an online format and split
into basic and advanced sessions.
Additionally, beginning with the AAS 238 online meeting, the workshops have been
expanded to Summer AAS meetings.
The Learn team finds that the workshop audience is best found at AAS meetings as
opposed to more general \python meetings, as the content tends to be more
applicable for students and researchers in astronomy and astrophysics.

The Astropy Project recently provided another mode of community engagement at
AAS Meetings 235 and 237 by organizing a NumFOCUS Sponsored Projects booth in
the AAS Exhibit Hall.
Funding for the exhibit hall was provided alternately by NumFOCUS and later by
the Moore Foundation funding.
The booth hosted a series of Q\&A special sessions during AAS 235 and webinars
during the virtual AAS 237 meetings, to provide the general astronomy community
information and access to experts on a variety of open-source astronomical
tools.

% \subsubsection{Learn vision for the future}

The focus of Learn Astropy over the coming months and years is to further
improve the interface for sharing educational content on the main Learn Astropy
website, and to facilitate the development of new content that highlights
\astropypkg and \astropy-affiliated package functionality.
However, we simultaneously also plan to solicit indexing and ingestion of
third-party tutorial series, to provide a unified interface for identifying
astronomy-specific educational content that demonstrates the functionality
available in the vast ecosystem of open-source software packages.
We also plan to look for opportunities to expand the reach of \astropy workshops
beyond the AAS meetings.

% \begin{itemize}
% \item {\bf User forums:} The Astropy Project has historically maintained a number of avenues for users to seek help or gain access to the developer community. This includes the Astropy mailing list, the Python in Astronomy Facebook group, and the Astropy Slack workspace. The Moore Foundation grant allows this Slack space to
% be on a paid plan; additionally NumFOCUS has negotiated a special rate for
% open-source projects. While the Slack and Astropy-dev mailing lists are primarily used to discuss the project direction and updates, it was noted that the use of a Facebook community could present a barrier to open source. The Astropy Project identified a need for a public, archived, searchable, and easily-to-navigate interface for users to ask for help Accordingly, we have commissioned
% a Discourse site which is more open than the Facebook group and more user-oriented
% than Astropy Slack. A benefit of the Moore Foundation grant is that Astropy
% developers are able to invoice as independent contractors the time they
% spend helping users on these forums.
% \end{itemize}

\section{Future Plans for the Astropy Project} \label{sec:future}

\subsection{A Roadmap for Future Priorities}
% \secauthor{Clara Brasseur}

The
\href{https://github.com/astropy/astropy-project/blob/main/roadmap/roadmap.md}{Astropy Roadmap} is a document in the \texttt{astropy-project} \github
repository within the \astropy organization that captures high-level actionable
items that the Astropy Project aims to undertake to improve the health and
stability of the Project. It is a static document that is revisited regularly at
the Astropy coordination meetings, intended to keep track of progress and write
new versions as needed. There is a related project board linked to the Roadmap
document that holds specific issues and efforts related to Roadmap items. The
project board is a living document that is continually updated as work is
planned, assigned, and completed.

All items in the Roadmap have been agreed to be priorities for the Astropy
Project, and are color-coded based on resources (both time/effort and
developers/expertise) needed to complete the item. Green items are well
underway, have sufficient resources/support and a plan in place for completion;
for example, efforts to improve the discoverability of documentation and
educational materials by overhauling the Learn website, which is underway (see
Section~\ref{sec:learn}).
Orange items are well defined, and work for acquiring sufficient resources
underway; for example, the goal of providing next-generation spectroscopic
reduction, analysis, and visualization tools usable by individual researchers
and larger surveys.
Red items do not yet have a plan for implementation and need more resources; for
example, the goal of improving interoperability with performant I/O file formats
and libraries such as HDF5 and Dask.
The Astropy Roadmap originates from the March 2021 Astropy Coordination meeting
where it was first drafted before being handed off to a newly formed Astropy
Roadmap working group for completion. The Roadmap in its current form was
adopted via pull request in December 2021.

\begin{figure}[th!]
  \begin{centering}
    \includegraphics[width=\textwidth]{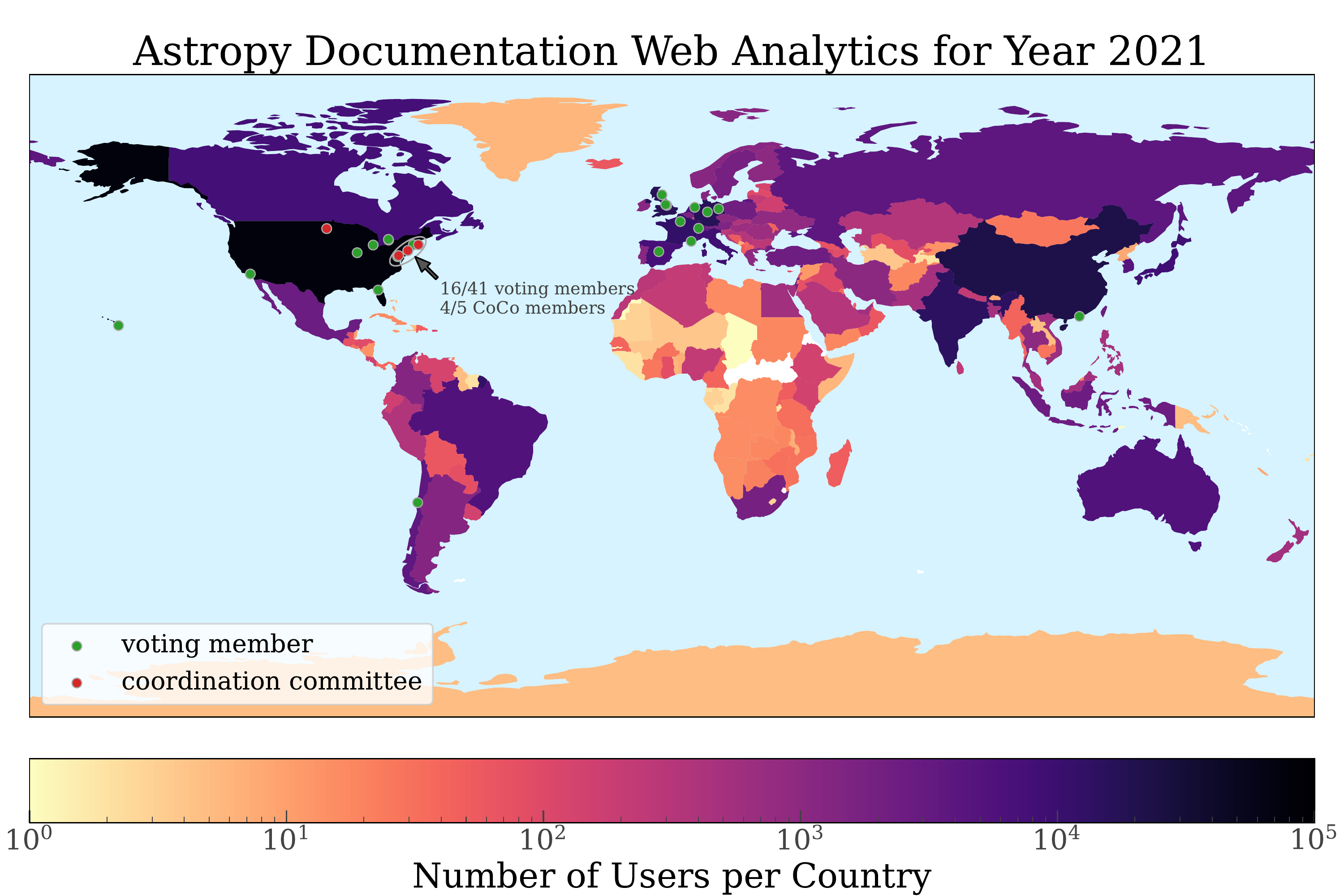}
      \caption{
          The number of users who accessed the \astropypkg core package
          documentation in 2021, based on data from Google Analytics.
          The circle markers show the approximate locations of home institutions
          for \astropy voting members (green) and coordination committee members
          (red), demonstrating the over-representation of the United States and,
          less so, Europe given the global spread of \astropypkg users.
      }
      \label{fig:User-analytics-locations}
  \end{centering}
\end{figure}

\subsection{Current and Future Challenges}

Over the history of \astropy, the needs and challenges of the \astropy Project
have evolved, but many common themes have persisted and are shared by other
open-source software communities.
However, \astropy also faces a number of unique challenges stemming from the
fact that the Project is in a special position in the astronomy software stack:
the \astropy code and ecosystem are more specialized than core scientific tools
like \package{numpy} and \package{scipy}, but more general than software
packages developed to enable research in specific subdomains such as
\package{exoplanet} \citep{exoplanet:2021}.
Below we discuss a few challenges faced by the Project in more detail, but we
stress that this list is not complete or exhaustive.
\begin{description}
  \item[Supporting an active group of maintainers] Throughout its history,
  development and maintenance of code and educational materials in the \astropy
  Project --- the \astropypkg core package, coordinated packages, Learn Astropy
  repositories, and Project infrastructure packages --- has been done by a mix
  of volunteer effort and institutional support.
  At present, the \href{https://www.astropy.org/team.html}{named roles} and
  maintainer positions in the \astropy Project are also split roughly evenly
  between these types of contributors.
  As is the case with many other open-source software communities, the culture
  of the Project and overall coordination relies heavily on volunteer effort.
  We therefore have significant amounts of critical infrastructure or core
  functionality that is either maintained with no redundancy or unmaintained as
  contributors naturally move on from and stop contributing to the Project.
  A prime example within the core package is the \astropyfits subpackage, which
  is largely in maintenance mode and maintained by just one contributor (despite
  the significant codebase and user community behind this subpackage).
  While we now have funding to support the longer-term sustainability of the
  Project and thus to address these issues, we are finding it difficult to use
  this funding in ways that address our needs.
  For one, many of the needs of the Project require long-term and sustained
  funding rather than short-term contracts.
  Furthermore, the specialized nature of the software requires being able to hire or
  contract software engineers with domain knowledge about astronomical software
  tools.
  Finally, we want to be able to preserve and sustain the engaged community of
  volunteers, software-engaged researchers, and existing software engineers that
  contribute to the Project and that have enabled its success in ways that
  support their careers.
  These challenges will require both cultural and larger systemic changes to
  astronomical research and career paths, and so will take time to influence and
  implement.
  \item[Diversity of users, contributors, maintainers, and coordinators]
  The \astropy Project aims to be made up of members at all levels (users to
  contributors to coordinators) with a diverse set of skills, backgrounds, and
  experiences.
  While \astropy is used globally --- for example,
  Figure~\ref{fig:User-analytics-locations} shows the number of users per
  country who accessed the \astropypkg core package documentation in 2021 ---
  our user community is underrepresented in non-English-speaking or -teaching
  countries, and our maintainers and coordinators are fairly narrowly
  concentrated in the United States (US) and Europe (see the circle markers on
  Figure~\ref{fig:User-analytics-locations}).
  The gap in users in many countries is understandable given that much of our
  content and documentation is written in English and therefore only available
  in other languages through imperfect, automated translation services (like
  Google translate).
  The fact that our maintainers and coordinators are generally in the US (even
  then, mostly on the East coast) and Europe is a reflection of where
  institutions have explicitly supported \astropy, and where volunteers have had
  the ability to contribute to the Project.
  \item[Long-term and sustained funding for maintaining infrastructure] The
  software and packaging infrastructure within the \astropy Project includes
  (but is not limited to) Continuous Integration (CI), test plugins,
  documentation builds, and package delivery.
  Keeping the infrastructure working is essential in software development and
  operations.
  This kind of work has low visibility and is not attractive to scientists or
  volunteer contributors; therefore, it is most effective to delegate such tasks
  to paid roles.
  \astropy secured a three-year grant for this but the situation remains
  uncertain once the grant runs out, as infrastructure evolves and requires
  constant maintenance and upgrades.
  Some of the engineers currently in these roles are also employees of research
  institutions that may leave the Project without much notice, adding to the
  uncertainty factor.
  \item[Understanding, communicating with, and engaging our user communities]
  The \astropy Project is used by a number of intersecting and overlapping
  communities, such as telescope operations and data management, data archives,
  researchers, students, educators, and the general public.
  These communities have different needs from and uses for \astropy, many of
  which we are only aware of through individual engagement from members of these
  respective communities.
  This kind of communication may provide biased perspectives on these
  communities.
  We therefore need to develop more formal and democratic ways of engaging these
  communities, such as through running and analyzing user surveys, or
  creating and engaging with user groups that can better represent these
  communities.
  Additionally, as the \astropy ecosystem has and continues to grow, we need to
  ensure that users in these different communities have access to educational
  materials to train new users to use \astropy and \python software tools.
  This may involve tailoring workshops, tutorials, and educational materials to
  these communities, however this represents a significant maintenance and
  content-development burden on the Project.
\end{description}

\section{Conclusions}
\label{sec:conclusions}

The Astropy Project has continued to serve as an active, community-oriented
software project with a wide reach and growing impact, both within astronomy and
astrophysics and in broader scientific software communities.
The Project is in the midst of setting up infrastructure to ensure its long-term
sustainability and to continue healthy growth of the software ecosystem,
developer, and user communities that interface with the Project.
For example, over the last years, we have formalized and implemented a new
governance model for the Astropy Project (Section~\ref{sec:project-governance}),
have continued to engage a growing pool of active and top contributors
(Section~\ref{sec:project-contributors}), and have begun to receive significant
formal funding for the Project itself (Section~\ref{sec:project-funding}).
The Astropy Project is also expanding its educational and community-focused
initiatives by soliciting and funding initiatives in inclusion, diversity, and
equity (Section~\ref{sec:project-ide}) as well as through the Learn Astropy initiative,
which aims to make tutorials and educational content that highlight software
tools for astronomy more discoverable and accessible (Section~\ref{sec:learn}).
We expect to continue in all of these threads over the coming years as we aim to
make the Astropy Project more sustainable and a larger, more welcoming community
for open-source software users and enthusiasts in astronomy.

% Due to a bug in AAStex 6.3.1, using this with amsmath leads to line numbers in
% the acknowledgements...
% https://github.com/AASJournals/AASTeX60/issues/130
% \begin{acknowledgments}
\vspace{2em}

We would like to thank the members of the community who have contributed to
\astropy, opened issues and provided feedback, and supported the
Project in a number of different ways.

We acknowledge the Gordon and Betty Moore foundation for their continued
financial support.
This work is partially supported by NASA under Grant No. 80NSSC22K0347 issued
through the NASA ROSES program.
This work is partially supported by the international Gemini Observatory, a
program of NSF's NOIRLab, which is managed by the Association of Universities
for Research in Astronomy (AURA) under a cooperative agreement with the National
Science Foundation, on behalf of the Gemini partnership of Argentina, Brazil,
Canada, Chile, the Republic of Korea, and the United States of America.
We also thank NumFOCUS and the Python Software Foundation for financial
support.

JAA-M acknowledges funding support from Macquarie University through the
International Macquarie University Research Excellence Scholarship (`iMQRES').
AB was supported by the Lend\"ulet Program of the Hungarian Academy of Sciences,
project No. LP2018-7, and the KKP-137523 `SeismoLab' \'Elvonal grant of the
Hungarian Research, Development and Innovation Office (NKFIH).
MB gratefully acknowledges support by the ANID BASAL project FB210003 and the
FONDECYT regular grant 1211000.
FDE acknowledges funding through the H2020 ERC Consolidator Grant 683184, the
ERC Advanced grant 695671 ``QUENCH'' and support by the Science and Technology
Facilities Council (STFC).
SG made contributions to Astropy a part of the Google Summer of Code 2021.
SG acknowledges some intellectual support from the CCAD-UNC-NOVA.
Funding for the Stellar Astrophysics Centre is provided by The Danish National
Research Foundation (Grant agreement no.: DNRF106).
NK acknowledges support from the MIT Pappalardo fellowship.
KAO acknowledges support by the European Research Council (ERC) through Advanced
Investigator grant to C.S. Frenk, DMIDAS (GA 786910).
CP is supported by the Canadian Space Agency under a contract with NRC Herzberg
Astronomy and Astrophysics.
SP has been supported by Spanish MINECO-FEDER grant RTI2018-096188-B-I00
JPG acknowledges funding support from Spanish public funds for research from
project PID2019-107061GB-C63 from the ``Programas Estatales de Generaci\'on de
Conocimiento y Fortalecimiento Cient\'ifico y Tecnol\'ogico del Sistema de I+D+i
y de I+D+i Orientada a los Retos de la Sociedad,'' as well as from the State
Agency for Research of the Spanish MCIU through the ``Center of Excellence
Severo Ochoa'' award to the Instituto de Astrof\'{\i}sica de Andaluc\'{\i}a
(SEV-2017-0709).
NS acknowledges support from the National Science Foundation through the
Graduate Research Fellowship Program under Grant 1842402. Any opinions,
findings, and conclusions or recommendations expressed in this material are
those of the authors and do not necessarily reflect the views of the National
Science Foundation.
Supported by the international Gemini Observatory, a program of NSF's NOIRLab,
which is managed by the Association of Universities for Research in Astronomy
(AURA) under a cooperative agreement with the National Science Foundation, on
behalf of the Gemini partnership of Argentina, Brazil, Canada, Chile, the
Republic of Korea, and the United States of America.
Parts of this research was supported by the Australian Research Council Centre
of Excellence for All Sky Astro-physics in 3 Dimensions (ASTRO 3D), through
project number CE170100013.
DS is supported by STFC grant ST/S000240/1.
NS acknowledges the support of the Science and Engineering Research Council of
Canada (NSERC) Canadian Graduate Scholarship — Doctoral Program, [funding
reference numbers CGSD — 54721 — 2020].

The \astropy community is supported by and makes use of a number of
organizations and services outside the traditional academic community.
We thank Google for financing and organizing the
Google Summer of Code (GSoC) program, that has funded severals
students per year to work on \astropy related projects over the
summer. These students often turn into long-term contributors.
Within the academic community, we thank institutions that make it possible for
astronomers and other developers on their staff to contribute their time to the
development of \astropy projects.
We would like acknowledge the support of the Space Telescope Science Institute,
Harvard--Smithsonian Center for Astrophysics, and the South African Astronomical
Observatory.

Furthermore, the \astropy packages would not exist in their current form without
a number of web services for code hosting, continuous integration, and
documentation; in particular, \astropy heavily relies on \github, Azure
Pipelines, CircleCI, and Read the Docs.

This research has made use of NASA's Astrophysics Data System.

\astropypkg interfaces with the SIMBAD database, operated at CDS, Strasbourg,
France. It also makes use of the \package{ERFA} library \citep{erfa}, which in turn
derives from the IAU SOFA Collection\footnote{\url{http://www.iausofa.org}}
developed by the International Astronomical Union Standards of Fundamental
Astronomy \citep{sofa}.

% \end{acknowledgments}

\software{\package{astropy} \citep{astropy:2013, astropy:2018},
          \package{Cython} \citep{cython}.
          \package{geopandas} \citep{geopandas},
          \package{matplotlib} \citep{matplotlib},
          \package{numpy} \citep{numpy:nature},
          \package{pandas} \citep{pandas, pandas2},
          \package{scipy} \citep{scipy}.
          % \package{photutils} \citep{photutils},
          % \package{specutils} \citep{specutils},
          % \package{regions} \citep{regions},
          % \package{gwcs} \citep{gwcs},
          % \package{synphot} \citep{synphot},
          % \package{ASDF} \citep{ASDF}.
          % \package{SOFA} \citep{sofa},
          % \package{ERFA} (\citealt{erfa})
          }

\bibliographystyle{aasjournal}
\bibliography{refs, static/affiliated-refs}

\appendix

\section{List of Affiliated Packages}

\begin{longrotatetable}
    \begin{deluxetable*}{ccp{2in}p{2.5in}}
    \tablecaption{Registry of affiliated packages.}
    \label{tab:affiliated-registry}
    \tablehead{
        \colhead{Package Name} &
        % \colhead{Stable} &
        \colhead{PyPI Name} &
        \colhead{Maintainer(s)} &
        \colhead{Citation(s)}
      }
      \startdata
        \href{https://github.com/cosimoNigro/agnpy}{agnpy} & \href{https://pypi.python.org/pypi/agnpy}{agnpy} & Cosimo~Nigro & \citet{agnpy} \\
\href{https://github.com/aplpy/aplpy}{APLpy} & \href{https://pypi.python.org/pypi/APLpy}{APLpy} & Thomas~Robitaille,\newline Eli~Bressert & \citet{APLpy:1},\newline \citet{APLpy:2} \\
\href{https://github.com/astropy/astroscrappy}{Astro-SCRAPPY} & \href{https://pypi.python.org/pypi/astroscrappy}{astroscrappy} & Curtis~McCully & \citet{Astro-SCRAPPY} \\
\href{https://github.com/quatrope/astroalign}{astroalign} & \href{https://pypi.python.org/pypi/astroalign}{astroalign} & Martin~Beroiz & \citet{astroalign} \\
\href{https://github.com/astroML/astroML}{astroML} & \href{https://pypi.python.org/pypi/astroML}{astroML} & Jake~Vanderplas,\newline Brigitta~Sip\H{o}cz & \citet{astroML} \\
\href{https://github.com/astropy/astroplan}{astroplan} & \href{https://pypi.python.org/pypi/astroplan}{astroplan} & Brett~Morris & \citet{astroplan} \\
\href{https://github.com/astropy/astropy-healpix}{astropy-healpix} & \href{https://pypi.python.org/pypi/astropy-healpix}{astropy-healpix} & Thomas~Robitaille,\newline Leo~Singer &  \\
\href{https://github.com/astropy/astroquery}{astroquery} & \href{https://pypi.python.org/pypi/astroquery}{astroquery} & Adam~Ginsburg,\newline Brigitta~Sip\H{o}cz & \citet{astroquery} \\
\href{https://github.com/mhvk/baseband}{baseband} & \href{https://pypi.python.org/pypi/baseband}{baseband} & Marten~H.~van~Kerkwijk & \citet{baseband} \\
\href{https://github.com/dokester/BayesicFitting}{BayesicFitting} & \href{https://pypi.python.org/pypi/BayesicFitting}{BayesicFitting} & Do~Kester,\newline Migo~Mueller & \citet{BayesicFitting} \\
\href{https://github.com/astropy/ccdproc}{ccdproc} & \href{https://pypi.python.org/pypi/ccdproc}{ccdproc} & Steven~Crawford,\newline Matt~Craig & \citet{ccdproc} \\
\href{https://github.com/jesford/cluster-lensing}{cluster-lensing} & \href{https://pypi.python.org/pypi/cluster-lensing}{cluster-lensing} & Jes~Ford & \citet{cluster-lensing} \\
\href{https://github.com/toros-astro/corral}{corral} & \href{https://pypi.python.org/pypi/corral-pipeline}{corral-pipeline} & Toros~Survey~Team & \citet{corral} \\
\href{https://github.com/karllark/dust_extinction}{dust\_extinction} & \href{https://pypi.python.org/pypi/dust\_extinction}{dust\_extinction} & Karl~Gordon &  \\
\href{https://github.com/einsteinpy/einsteinpy}{EinsteinPy} & \href{https://pypi.python.org/pypi/einsteinpy}{einsteinpy} & Shreyas~Bapat,\newline Ritwik~Saha,\newline Bhavya~Bhatt,\newline Priyanshu~Khandelwal & \citet{EinsteinPy} \\
\href{https://github.com/carpyncho/feets}{feets} & \href{https://pypi.python.org/pypi/feets}{feets} & Juan~B~Cabral & \citet{feets} \\
\href{https://github.com/adrn/gala}{gala} & \href{https://pypi.python.org/pypi/gala}{gala} & Adrian~Price-Whelan & \citet{gala} \\
\href{https://github.com/jobovy/galpy}{galpy} & \href{https://pypi.python.org/pypi/galpy}{galpy} & Jo~Bovy & \citet{galpy} \\
\href{https://github.com/gammapy/gammapy}{gammapy} & \href{https://pypi.python.org/pypi/gammapy}{gammapy} & Axel~Donath,\newline R\'egis~Terrier & \citet{gammapy} \\
\href{https://github.com/ejeschke/ginga}{ginga} & \href{https://pypi.python.org/pypi/ginga}{ginga} & Eric~Jeschke,\newline Pey-Lian~Lim & \citet{ginga} \\
\href{https://github.com/glue-viz/glue}{Glue} & \href{https://pypi.python.org/pypi/glueviz}{glueviz} & Chris~Beaumont,\newline Thomas~Robitaille & \citet{Glue:1},\newline \citet{Glue:2} \\
\href{https://github.com/spacetelescope/gwcs}{gwcs} & \href{https://pypi.python.org/pypi/gwcs}{gwcs} & Nadia~Dencheva & \citet{gwcs} \\
\href{https://github.com/astropy/halotools}{Halotools} & \href{https://pypi.python.org/pypi/halotools}{halotools} & Andrew~Hearin &  \\
\href{https://github.com/StingraySoftware/HENDRICS}{HENDRICS} & \href{https://pypi.python.org/pypi/hendrics}{hendrics} & Matteo~Bachetti & \citet{HENDRICS} \\
\href{https://github.com/hipspy/hips}{hips} & \href{https://pypi.python.org/pypi/hips}{hips} & Thomas~Boch &  \\
\href{https://github.com/spacetelescope/imexam}{imexam} & \href{https://pypi.python.org/pypi/imexam}{imexam} & Megan~Sosey & \citet{imexam} \\
\href{https://github.com/legau/kanon}{kanon} & \href{https://pypi.python.org/pypi/kanon}{kanon} & L\'eni~Gauffier & \citet{kanon} \\
\href{https://github.com/sibirrer/lenstronomy}{lenstronomy} & \href{https://pypi.python.org/pypi/lenstronomy}{lenstronomy} & Simon~Birrer & \citet{lenstronomy:1},\newline \citet{lenstronomy:2} \\
\href{https://git.ligo.org/lscsoft/ligo.skymap}{ligo.skymap} & \href{https://pypi.python.org/pypi/ligo.skymap}{ligo.skymap} & Leo~Singer &  \\
\href{https://github.com/linetools/linetools}{linetools} & \href{https://pypi.python.org/pypi/linetools}{linetools} & J.~Xavier~Prochaska,\newline Nicolas~Tejos,\newline Neil~Crighton & \citet{linetools} \\
\href{https://github.com/Chandra-MARX/marxs}{marxs} & \href{https://pypi.python.org/pypi/marxs}{marxs} & Hans~Moritz~G\"unther & \citet{marxs} \\
\href{https://github.com/cds-astro/mocpy}{mocpy} & \href{https://pypi.python.org/pypi/mocpy}{mocpy} & Matthieu~Baumann,\newline Thomas~Boch &  \\
\href{https://github.com/zblz/naima}{naima} & \href{https://pypi.python.org/pypi/naima}{naima} & Victor~Zabalza & \citet{naima} \\
\href{https://github.com/RiceMunk/omnifit}{omnifit} & \href{https://pypi.python.org/pypi/omnifit}{omnifit} & Aleksi~Suutarinen & \citet{omnifit} \\
\href{https://github.com/astropy/photutils}{photutils} & \href{https://pypi.python.org/pypi/photutils}{photutils} & Larry~Bradley,\newline Brigitta~Sip\H{o}cz & \citet{photutils} \\
\href{https://github.com/poliastro/poliastro}{poliastro} & \href{https://pypi.python.org/pypi/poliastro}{poliastro} & Juan~Luis~Cano~Rodr{\'\i}guez & \citet{poliastro} \\
\href{https://github.com/gwastro/pycbc}{PyCBC} & \href{https://pypi.python.org/pypi/pycbc}{pycbc} & Alexander~Harvey~Nitz & \citet{PyCBC} \\
\href{https://github.com/weaverba137/pydl}{PyDL} & \href{https://pypi.python.org/pypi/pydl}{pydl} & Benjamin~Alan~Weaver & \citet{PyDL} \\
\href{https://github.com/astropy/pyregion}{pyregion} & \href{https://pypi.python.org/pypi/pyregion}{pyregion} & Jae-Joon~Lee &  \\
\href{https://github.com/pyspeckit/pyspeckit}{pyspeckit} & \href{https://pypi.python.org/pypi/pyspeckit}{pyspeckit} & Adam~Ginsburg & \citet{pyspeckit} \\
\href{https://github.com/astropy/pyvo}{PyVO} & \href{https://pypi.python.org/pypi/pyvo}{pyvo} & Christine~Banek,\newline Adrian~Demian,\newline Stefan~Becker & \citet{PyVO} \\
\href{https://github.com/astropy/regions}{regions} & \href{https://pypi.python.org/pypi/regions}{regions} & Larry~Bradley,\newline Adam~Ginsburg & \citet{regions} \\
\href{https://github.com/astropy/reproject}{reproject} & \href{https://pypi.python.org/pypi/reproject}{reproject} & Thomas~Robitaille &  \\
\href{https://github.com/astropy/saba}{Saba} & \href{https://pypi.python.org/pypi/saba}{saba} & Michele~Costa &  \\
\href{https://github.com/NASA-Planetary-Science/sbpy}{sbpy} & \href{https://pypi.python.org/pypi/sbpy}{sbpy} & Michael~Mommert,\newline Michael~S.~P.~Kelley,\newline Miguel~de~Val-Borro,\newline Jian-Yang~Li & \citet{sbpy} \\
\href{https://github.com/sncosmo/sncosmo}{sncosmo} & \href{https://pypi.python.org/pypi/sncosmo}{sncosmo} & Kyle~Barbary & \citet{sncosmo} \\
\href{https://github.com/radio-astro-tools/spectral-cube}{spectral-cube} & \href{https://pypi.python.org/pypi/spectral-cube}{spectral-cube} & Adam~Ginsburg & \citet{spectral-cube} \\
\href{https://gitlab.com/mcfuman/SpectraPy/}{SpectraPy} & \href{https://pypi.python.org/pypi/}{} & Marco~Fumana & \citet{SpectraPy} \\
\href{https://github.com/astropy/specutils}{specutils} & \href{https://pypi.python.org/pypi/specutils}{specutils} & Nicholas~Earl,\newline Adam~Ginsburg,\newline Erik~Tollerud &  \\
\href{https://github.com/spacetelescope/spherical_geometry}{spherical\_geometry} & \href{https://pypi.python.org/pypi/spherical-geometry}{spherical-geometry} & Bernie~Simon &  \\
\href{https://github.com/vrodgom/statmorph}{statmorph} & \href{https://pypi.python.org/pypi/statmorph}{statmorph} & Vicente~Rodriguez-Gomez & \citet{statmorph} \\
\href{https://github.com/StingraySoftware/stingray}{stingray} & \href{https://pypi.python.org/pypi/stingray}{stingray} & Daniela~Huppenkothen,\newline Matteo~Bachetti,\newline Abigail~Stevens,\newline Simone~Migliari,\newline Paul~Balm & \citet{stingray:1},\newline \citet{stingray:2} \\
\href{https://github.com/spacetelescope/synphot_refactor}{synphot} & \href{https://pypi.python.org/pypi/synphot}{synphot} & Pey~Lian~Lim & \citet{synphot} \\

      \enddata
  \end{deluxetable*}
\end{longrotatetable}

\end{document}